\pdfoutput=1
\documentclass[aps,pra,amsmath,amssymb,twocolumn,superscriptaddress,showpacs]{revtex4-1}

\usepackage{xcolor}
\usepackage{amsmath,amssymb}
\usepackage{graphicx}
\usepackage{ulem}
\usepackage{textcomp}

\def\vecb{\boldsymbol}

\begin{document}

\preprint{APS/123-QED}

\title{Role of electron scattering on the high-order harmonic generation from solids}

\author{Chang-Ming~Wang}

\affiliation 
{Max Planck Institute for the Structure and Dynamics of Matter, Luruper Chaussee 149, 22761 Hamburg, Germany}

\author{Nicolas~Tancogne-Dejean}
\affiliation 
{Max Planck Institute for the Structure and Dynamics of Matter, Luruper Chaussee 149, 22761 Hamburg, Germany}

\author{Massimo~Altarelli}
\affiliation 
{Max Planck Institute for the Structure and Dynamics of Matter, Luruper Chaussee 149, 22761 Hamburg, Germany}

\author{Angel~Rubio}
\email{angel.rubio@mpsd.mpg.de}
\affiliation 
{Max Planck Institute for the Structure and Dynamics of Matter, Luruper Chaussee 149, 22761 Hamburg, Germany}

\author{Shunsuke~A.~Sato}
\email{ssato@ccs.tsukuba.ac.jp}
\affiliation 
{Center for Computational Sciences, University of Tsukuba, Tsukuba, Ibaraki 305-8577, Japan}
\affiliation 
{Max Planck Institute for the Structure and Dynamics of Matter, Luruper Chaussee 149, 22761 Hamburg, Germany}

\date{\today}

%
%

\begin{abstract}
We extend the semi-classical trajectory description for the high-order harmonic generation~(HHG)
from solids by integrating the effect of electron-scattering.
Comparing the extended semi-classical trajectory model with a one-dimensional quantum mechanical
simulation, we find that the multi-plateau feature of the HHG spectrum is formed by Umklapp scattering under the electron-hole acceleration dynamics by laser fields.
Furthermore, by tracing the scattered trajectories in real-space, the model fairly describes the emitted photon energy
and the emission timing of the HHG even in the higher plateau regions.
We further consider the loss of trajectories by scattering processes with
a mean-free-path approximation and evaluate the HHG cutoff energy as
a function of laser wavelength. As a result, we find that the trajectory loss by scattering
causes the wavelength independence of the HHG from solids.
\end{abstract}

\maketitle


\section{Introduction \label{sec:intro}}

Light-matter interactions have been an important subject in physics from both fundamental
and technological points of view \cite{MOUROU2012720,Krausz_2014_Attosecond,Basov2017,Flick3026}.
Intense light may couple with matter nonlinearly and can induce nonlinear optical effects
\cite{butcher1990elements,2008v,Hentschel_2001_Attosecond,Pfeifer_2006_Femtosecond,Krausz_2009_Attosecond}, such as the second order harmonic generation
\cite{PhysRevLett.7.118}.
Once field strength of light becomes extremely large, non-perturbative and highly-nonlinear
phenomena may be induced. A primary example of such phenomena
is the high-order harmonic generation~(HHG) \cite{Krause_1992_High,Schafer_1993_Above,Lewenstein_1994_Theory}, which is an extreme
photon-upconversion process via strongly-nonlinear light-matter interactions.
This process has been observed from atomic gases decades ago \cite{McPherson_1987_Studies,Ferray_1988_Multiple},
and the gas-phase HHG further opened a novel
technology to generate ultrashort laser pulses with attosecond duration,
offering a novel avenue to explore ultrafast real-time electron dynamics in matter
\cite{Itatani2004,Goulielmakis2010,Schultze1348,Lucchini916,Volkov2019,Siegrist2019}.
Recently, the HHG from solid-state materials has been systematically observed
\cite{Ghimire2011}. It has been demonstrated that the solid-state HHG
shows distinct fundamental features from the gas-phase HHG, such as the linear scaling
of the cutoff energy with respect to the field strength \cite{Ghimire2011,Schubert_2014_Sub,Luu2015},
the wavelength independence of the cutoff energy
\cite{Ghimire2011,Luu2015,Nicolas_2017_Impact}, and the enhancement of HHG by elliptically-polarized light \cite{Yoshikawa736,Tancogne-Dejean2017}.
It has also been shown that the competition of mechanisms between atomic-like and solid-like responses in two-dimensional systems further enriches the HHG spectra \cite{Liu_2017_High,Nicolas_2018_Atomic,Guillaume_2018_High}.
These distinct features of the solid-state HHG have been drawing great attention
because it offers a novel possibility to investigate ultrafast electron dynamics
in matter and may open a path to novel light sources \cite{Ghimire2019}.

The mechanism of the HHG from gases has been understood by the semi-classical
trajectory model;
so-called \textit{three-step model}
\cite{kulander1993dynamics,Corkum_1994_Plasma,PhysRevA.49.2117}.
The model consists of the following three steps: (i) An electron is ionized from
an atom or molecule due to a strong laser field. (ii) The ionized electron is accelerated by
the laser field in vacuum. (iii) The accelerated high-energy electron returns to
the parent ion due to the oscillatory field and recombines,
emitting high energy photons. The three-step model describes well the features of the gas-phase HHG such as the cutoff energy.
The semi-classical trajectory model has been
further extended to the solid-state HHG \cite{Vampa_2015_semiclassical},
integrating the electronic band dispersion of solids as the crystal-momentum-dependent
effective mass of electron-hole pairs. In this regard,
the extended semi-classical trajectory model
still treats the dynamics of electron-hole pairs as that of free particles.
However, in contrast to ionized electrons in vacuum,
electrons in solids may be easily scattered by ions, other electrons, defects and so on.
Therefore, electron scattering is expected to play an important role for
the HHG from solids.

Despite the great effort to study the mechanism of the HHG from solids,
the role of the electron scattering has not been investigated yet in the context of the semi-classical trajectory description.
In this paper we consider an extension of the real-space semi-classical trajectory model 
by incorporating scattering effects in solids. The generalization is carried out
by branching a classical trajectory into multiple trajectories whenever a scattering event
occurs. We compare the scattering-integrated semi-classical trajectory model with
one-dimensional quantum dynamical simulations,
and explore a role of Umklapp scattering in the HHG from solids.
Furthermore, we extend our modeling with the mean-free-path approximation
and elucidate the effect of the scattering to the wavelength scaling of the HHG cutoff.

This paper is organized as follows: In Sec.~\ref{sec:methods} we first revisit
the semi-classical trajectory model for HHG from solids. Then, we extend the model
by incorporating the scattering effect.
In Sec.~\ref{sec:result} we examine the role of the scattering in the HHG with
the model, comparing with the one-dimensional quantum model.
Finally our findings are summarized in Sec.~\ref{sec:summary}.
In this work, atomic units are used unless stated otherwise.

\section{Methods \label{sec:methods}}
\subsection{Semi-classical trajectory model for solid-state HHG \label{subsec:classical_model}}

Here, we first revisit a semi-classical trajectory model for the solid-state HHG
\cite{Vampa_2015_semiclassical} to further introduce the scattering effect into
it. The semi-classical trajectory model, or the so-called \textit{three-step model}, has been
originally proposed to describe the HHG from noble gases
\cite{kulander1993dynamics,Corkum_1994_Plasma,PhysRevA.49.2117}. The model properly describes key features
of the HHG spectrum such as the cutoff energy. Recently, the semi-classical trajectory
model was extended to the solid-state HHG \cite{Vampa_2015_semiclassical}. 
In the solid-state semi-classical trajectory model, the HHG is described by the following three steps:
\begin{enumerate}
\item Creation of an electron-hole pair by exciting an electron from a valence band
to a conduction band at the optical gap of the solid.

\item Acceleration of the electron-hole pair by an external laser field. 

\item Emission of a high-energy photon by the recombination of the accelerated
electron-hole pair.

\end{enumerate}

The major difference of the solid-state semi-classical trajectory model from the corresponding gas-phase
model is the treatment of the acceleration of particles in the second step:
since ionized electrons travel in vacuum in the gas-phase model, they can be treated as
free charged-particles, which have a parabolic energy dispersion,
$\epsilon(\vecb k)=\hbar ^2 \vecb k^2/2m_e$ with the wavenumber $\vecb k$.
In contrast, the dynamics of electron-hole pairs in solids
is not generally described by the simple parabolic energy dispersion but requires
more complex anharmonic dispersion reflecting the solid-state electronic band structure.
Thus, the solid-state semi-classical trajectory model can be seen as
a generalization of the gas-phase model by changing the electron mass $m_e$ to
the effective electron-hole mass,
\begin{align}
\mu_{ij} = \left [ \frac{\partial^2}{\partial k_i \partial k_j} \left ( 
\epsilon_{c\vecb k} - \epsilon_{v\vecb k} 
\right ) \right ]^{-1},
\end{align}
where $\epsilon_{b \vecb k}$ is the band dispersion of valence ($b=v$) and
conduction ($b=c$) bands.

Based on the semi-classical trajectory model, the relative position $\vecb x(t)$ of
an electron-hole pair created at the time $t_0$ can be described as
\cite{Vampa_2015_semiclassical}:
\begin{align}
  \vecb x(t) = \int^t_{t_0}dt' \vecb v(\vecb K(t')),
    \label{e:classical_trajectory}
\end{align}
where $\vecb K(t)=\vecb k +\vecb A(t)$ is the shifted wavevector due to the applied
vector potential $\vecb A(t)$ by the acceleration theorem. The relative electron-hole
velocity $\vecb v(\vecb k)$ is determined by the electron-hole energy dispersion as
\begin{align}
  \vecb v(\vecb k)=\vecb v_c(\vecb k)- \vecb v_v(\vecb k) 
  =\frac{\partial}{\partial \vecb k}\left [
    \epsilon_{c\vecb k} - \epsilon_{v \vecb k}
    \right ].
\end{align}
In the semi-classical trajectory model, an electron-hole pair is assumed to be created at
the band gap with zero distance $\vecb x(t_0)=0$. Then, the trajectory $\vecb x(t)$
is evolved with Eq.~(\ref{e:classical_trajectory}). In the final step,
the electron-hole pair is recombined at time $t_r$ and emits
a photon when the electron and hole come back to the same position $\vecb x(t_r)=0$.
The emitted photon energy corresponds to the energy of the recombined electron-hole pair,
$\epsilon_{c \vecb K(t_r)}-\epsilon_{v \vecb K(t_r)}$.

\subsection{Electron scattering effect in the semi-classical trajectory model
  \label{subsec:scattering}}

In the above semi-classical model, dynamics of ionized electrons or created electron-hole
pairs is treated as independent free particles. This treatment is accurate enough to
describe the HHG from dilute gases. However, in solids, electrons and holes can be
scattered by phonons, other electrons and holes, impurities and many other processes.
Therefore, the free-particle treatment without scattering processes is not complete
to describe the solid-state HHG.
The importance of such scattering processes could also be seen in the recently observed multi-plateau feature in HHG spectra.
Unlike many HHG spectra from semiconductors featuring single plateau, the experiments using noble-gas solids \cite{Ndabashimiye_2016_Solid} demonstrate the multiple plateaus feature
in the HHG specta.
Furthermore, it has been demonstrated that such multi-plateau feature can be theoretically described by ladder-climbing process \cite{Ikemachi_2017_Trajectory}, which is conceptually based on Umklapp scattering. 
Thus, the multi-plateau feature could be seen as the consequence of the scattering effect. 
However, in the previous study proposing the ladder-climbing process, only the $k$-space (crystal momentum space) semi-classical trajectories have been considered,
and it was assumed that electron-hole pairs can recombine at any instance of time. As a result, an electron and a hole are allowed to recombine and emit a photon no matter how far
they are separated in real space.
In this work, we integrate Umklapp scattering effect into the semi-classical trajectory model using both real- and $k$-space trajectories.
This allows us to obtain the information of recombination time for electron-hole pairs and thereby refine the solid-state semi-classical trajectory model.

\begin{figure}[htbp]
\includegraphics[width=\columnwidth]{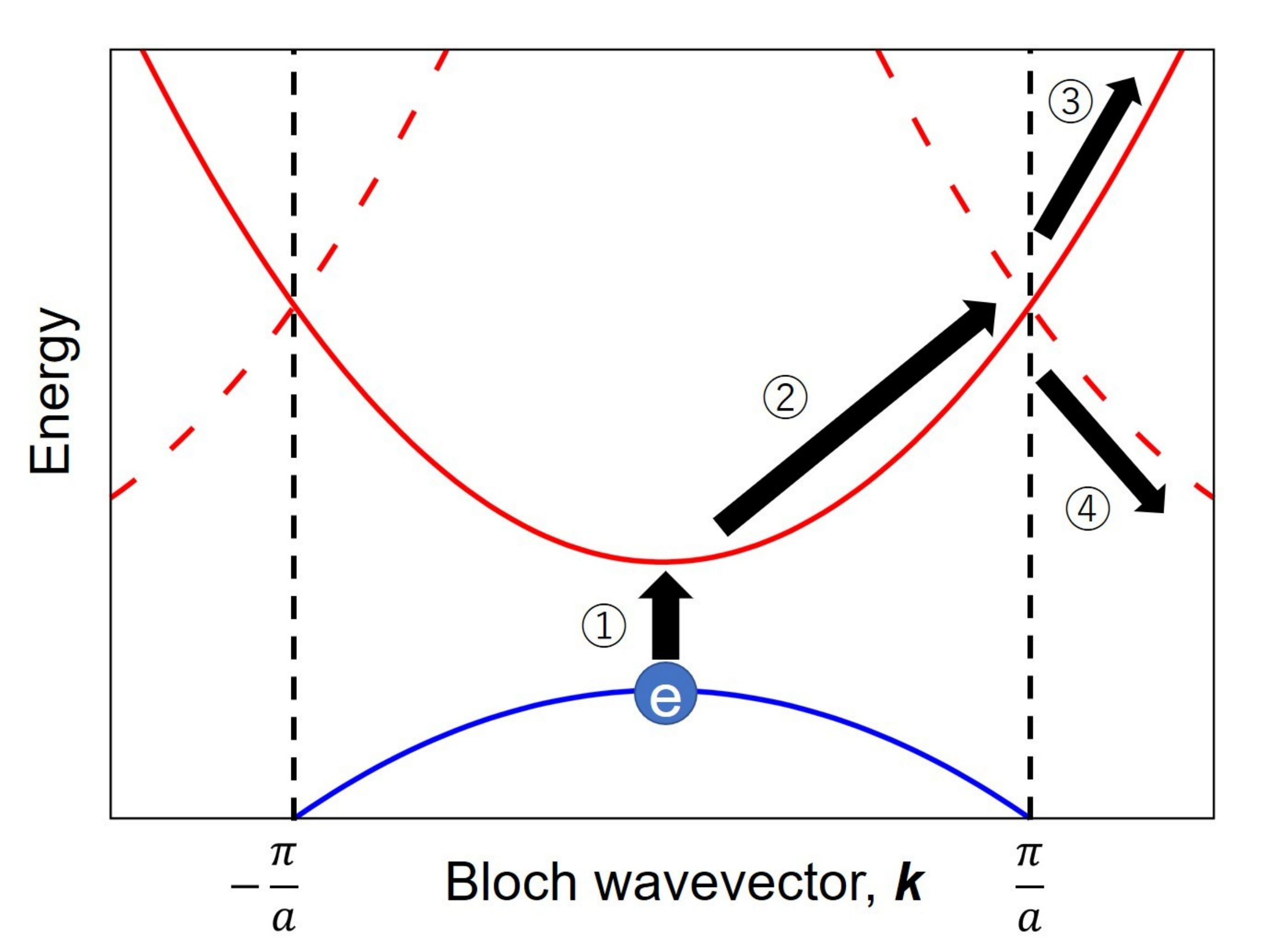}
\caption{\label{fig:schematic}
Schematic picture of semi-classical trajectory dynamics
and Umklapp scattering. The blue-solid line describes a valence band while the red-solid line
describes an unfolded conduction band. The replicated unfolded-bands with the reciprocal
lattice vector are indicated as the red-dashed lines.
}
\end{figure}

The incorporation of Umklapp scattering effect with the semi-classical trajectory model is carried out as follows:
\begin{enumerate}
\item Creating an electron-hole pair at the time $t_0$ by exciting
an electron from a valence band to a conduction band at the Bloch wavevector,
$\vecb k_0$ that corresponds to the optical gap of the solid.

\item \label{step:traj-scatter-02} Propagating the trajectory in real-space,
$\vecb x(t)$ with
Eq.~(\ref{e:classical_trajectory}) and the trajectory in $k$-space,
$\vecb k_0 +\vecb A(t)-\vecb A(t_0)$, with the acceleration theorem.

\item Branching a trajectory into scattered and non-scattered trajectories when
the trajectory reaches the Brillouin zone edge. Here, Umklapp scattering
is described as the sudden shift of the accelerated Bloch vector
$\vecb K(t) \rightarrow \vecb K(t) \pm n \vecb b$ with the reciprocal lattice vector,
$\vecb b$ and an integer $n$. The details of the scattering is depicted
in Fig.~\ref{fig:schematic}.
After the scattering, the branched trajectories are propagated in the same way described in the step~\ref{step:traj-scatter-02}.

\item Recombining the electron-hole pair at time $t_r$ when $\vecb x(t_r)=0$,
and emitting a photon whose energy corresponds to the energy difference
of the electron-hole pair, $\epsilon_{c,\vecb K(t_r)}-\epsilon_{v,\vecb K(t_r)}$.
\end{enumerate}

Figure~\ref{fig:schematic} shows a schematic picture of the branching procedure discussed above. The red-solid line shows an unfolded conduction band while
the blue-solid line shows a valence band. The replicated conduction bands shifted
with the reciprocal lattice vector are described as the red-dashed lines.
In Fig.~\ref{fig:schematic}, the arrow, \textcircled{\footnotesize 1},
describes the excitation of an electron from the top of the valence band to the bottom of the conduction band.
The arrow, \textcircled{\footnotesize 2}, indicates the acceleration of
the electron from the conduction bottom to the Brillouin zone edge.
At the Brillouin zone edge, the conduction bands have a crossing, or they
may have an avoid-crossing. After the crossing at the Brillouin zone edge,
the electron trajectory may diabatically follow the red-solid line as depicted
with the arrow, \textcircled{\footnotesize 3}, or may be switched to
the red-dashed line as \textcircled{\footnotesize 4}.
Since the red-dashed lines are nothing but the replicated bands with the crystal momentum shift
by the reciprocal lattice vector, the switching of the bands is nothing
but Umklapp scattering at the Brillouin zone edge.
In general, where bands are crossing, the inter-band transition may occur due to scattering processes. In this work, such scattering effect is incorporated
with switching of the trajectory on energy bands.

\section{Results \label{sec:result}}

In this section, we first examine the difference between the simple parabolic band dispersion
and the non-parabolic solid-state band dispersion in the context of
the semi-classical trajectory analysis. For this purpose, we employ
the Kane band model \cite{Kane_1957_Band}.
Then, we elucidate the role of Umklapp scattering in the HHG by comparing
the semi-classical trajectory model with the one-dimensional quantum electron
dynamics simulation. Finally, we further explore the effect of more general
scattering with the mean-free path approximation.

\subsection{Solid-state trajectory model without scattering \label{subsec:trajectory}}

To study the solid-state HHG with the semi-classical model, we first examine
the semi-classical trajectory analysis with the Kane band model \cite{Kane_1957_Band}
without the scattering contribution.
The Kane band model is widely used to model
the solid-state electronic band structures. For example,
the seminal work for field-induced ionization by Keldysh \cite{keldysh1965},
which finds applications in many fields
\cite{balling2013femtosecond,PhysRevLett.82.3883,doi:10.1117/12.2244833},
is based on the Kane band.

Here, we briefly assess the difference between the parabolic band dispersion
and the solid-state Kane band dispersion
in the semi-classical trajectory model of HHG.
The parabolic energy dispersion is described as
\begin{align}
\epsilon_{parabolic}(\vecb k) = \epsilon_g + \frac{|\vecb k|^2}{2\mu},
\end{align}
where $\epsilon_g$ is the band gap, and $\mu$ is the electron-hole reduced mass.
With the same parameters, the Kane band can be described as
\begin{align}
    \epsilon_{Kane}(\vecb k) = \epsilon_g
\sqrt{1+\frac{|\vecb k|^2}{\mu\epsilon_g}}.
\label{eq:kane_band}
\end{align}

\begin{figure}[htbp]
\includegraphics[width=\columnwidth]{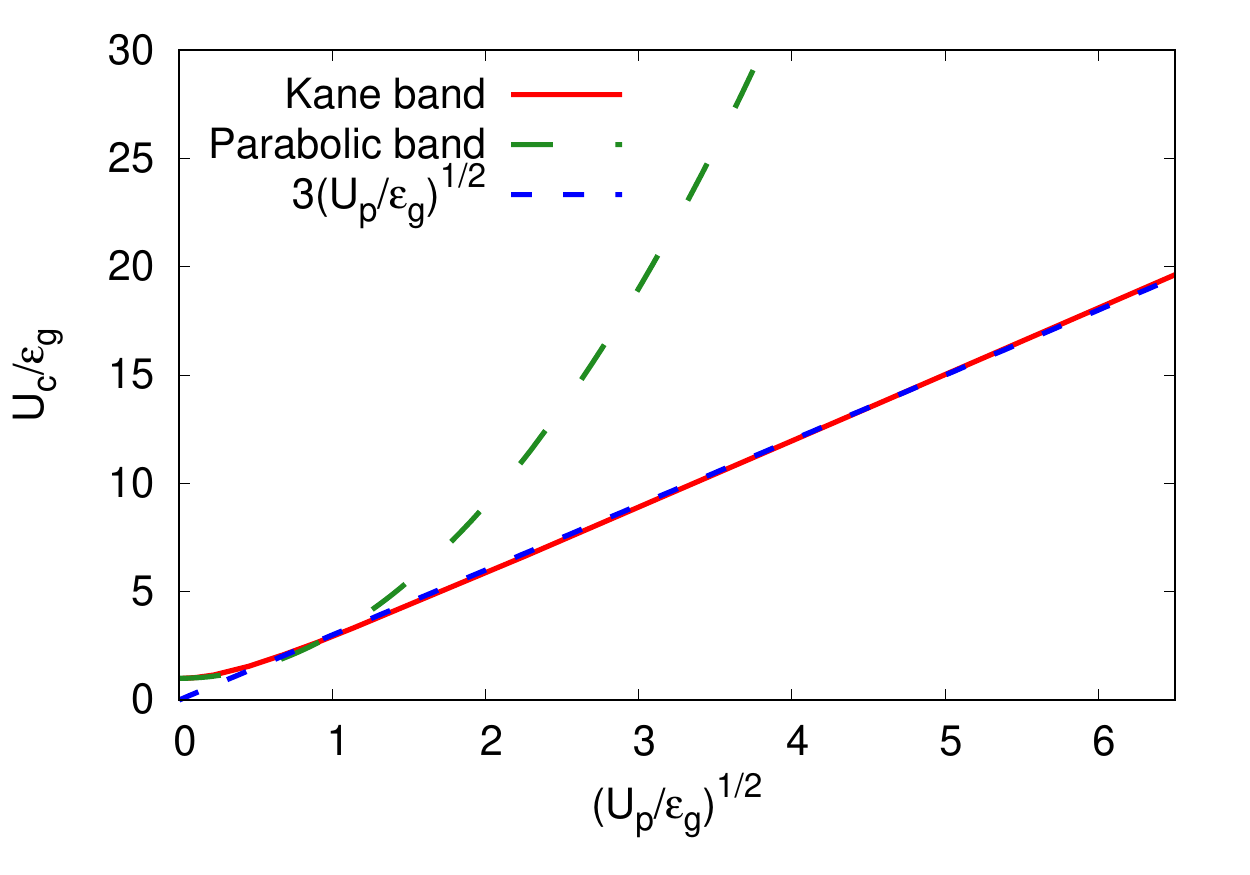}
\caption{\label{fig:cutoff} Cutoff energy of the HHG $U_c$ as a function of
applied laser field strength $F_0$, which is proportional to
the square root of the ponderomotive energy $U_p$.
The red-solid line shows the result of the Kane band model while the green-dashed line
shows that of the parabolic band model. The blue-dotted line indicates
the analytic line, $U_c/\epsilon_g=3(U_p/\epsilon_g)^{1/2}$.
}
\end{figure}

According to previous works \cite{PhysRevA.49.2117,Vampa_2015_semiclassical}
based on the semi-classical trajectory model, one can evaluate the cutoff energy
of HHG as the maximum recombination energy among all possible trajectories
under monochromatic laser fields.
Figure~\ref{fig:cutoff} shows the cutoff energy $U_c$ of the semi-classical
trajectory model as a function of the square
root of the ponderomotive energy $\sqrt{U_p}$, which is defined
as $U_p=F^2_0/4\mu \omega^2_0$ with the laser field strength $F_0$, the effective mass $\mu$
and the laser angular frequency $\omega_0$. The red-solid line shows
the result of the Kane band model, while the green-dashed line shows
that of the parabolic band model.
For the parabolic dispersion case, the cutoff energy $U_c$ is described
by the well-known formula as
\cite{PhysRevA.49.2117}
\begin{align}
U^{(parabolic)}_c = \epsilon_g + 3.17 U_p.
\end{align}

As seen from Fig.~\ref{fig:cutoff}, in the weak field region (small $U_p$ region),
the two models give the similar cutoff energy.
This observation can be understood by a fact that the Kane band model
is reduced to the parabolic band model in the small wavenumber limit as
$\epsilon_{Kane}(\vecb k)\rightarrow \epsilon_g +  |\vecb k|^2/2\mu, (|\vecb k|\rightarrow 0)$.
As seen from Fig.~\ref{fig:cutoff}, the two models show qualitatively different behaviors
in the strong field region: the cutoff energy of the Kane band model is proportional to
the field strength, $F_0\sim \sqrt{U_p}$, while that of the parabolic band model is
proportional to the square of the field strength. The quadratic field-strength dependence
of the parabolic band model is consistent with the cutoff energy of
the gas-phase HHG while
the linear field-strength dependence is consistent with the reported feature of
the solid-state HHG \cite{Ghimire2011,Schubert_2014_Sub,Luu2015}.

In Fig.~\ref{fig:cutoff}, the blue-dotted line shows the analytic line, 
$U_c/\epsilon_g = 3(U_p/\epsilon_g)^{1/2}$. One sees that the analytic blue-dotted line
shows nice agreement with the result of the Kane band model (red-solid line).
Therefore, in the strong field limit, the cutoff energy of the Kane band model is well
approximated as
\begin{align}
U^{Kane}_c \approx 3 \epsilon_g \sqrt{U_p/\epsilon_g}
= \frac{3}{2}v^{\infty}_g \frac{F_0}{\omega_0},
\label{eq:cutoff_kane}
\end{align}
where $v^{\infty}_g$ is the group electron-hole velocity of the Kane band model
in the large $|\vecb k|$ limit,
\begin{align}
v^{\infty}_g = \lim_{|\vecb k|\rightarrow \infty}\left | 
\frac{\partial}{\partial \vecb k}\epsilon_{Kane}(\vecb k)
\right | = \sqrt{\frac{\epsilon_g}{\mu}}.
\end{align}

In contrast, the $k$-space semi-classical trajectory model \cite{Ikemachi_2017_Trajectory}
provides the following cutoff formula in the strong field limit
\begin{align}
U^{k-\mathrm{space}}_c = \epsilon_{Kane} \left( \frac{2F_0}{\omega_0} \right)
\approx 2 \sqrt{\frac{\epsilon_g}{\mu}} \frac{F_0}{\omega_0}
= 2 v^{\infty}_g \frac{F_0}{\omega_0}.
\label{eq:cutoff_kane_kspace}
\end{align}
Note that the same cutoff expression has been derived for
the analysis on graphene \cite{Chizhova_2017_High-harmonic} with a similar consideration
to the $k$-space trajectory.
Comparing Eq.~(\ref{eq:cutoff_kane}) and Eq.~(\ref{eq:cutoff_kane_kspace}),
the real-space trajectory model provides the smaller cutoff energy
than the $k$-space trajectory model by a factor of $3/4$.
In the $k$-space trajectory model, the recombination is allowed at any instance of time.
On the other hand, in the real-space trajectory model, the recombination is allowed
only when paired electron and hole come to the same position. Therefore,
the recombination events in the real-space trajectory model is a subset of
those in the $k$-space model. Hence the cutoff energy of the real-space model
is smaller than that of the $k$-space model by construction.

\subsection{Umklapp scattering contribution to the HHG  \label{subsec:result_Umklapp_scatter}}

Here, we investigate the role of Umklapp scattering in the semi-classical trajectory
model. For this purpose, we compare the semi-classical trajectory model with
the one-dimensional quantum mechanical (1D-QM) simulation.
We employ the same model for the 1D-QM simulation as the previous work
\cite{Ikemachi_2017_Trajectory}.
We first briefly explain the 1D-QM simulation, and then introduce
the corresponding semi-classical trajectory model with Umklapp scattering.

The 1D-QM system is described by the following single-particle Schr\"odinger equation
\begin{align}
    i \frac{\partial u_{b,k}(x,t)}{\partial t}=
\left[\frac{1}{2} \left \{ -i\frac{\partial}{\partial x} + k + A(t) \right \}^2 
  + v(x)\right ] u_{b,k}(x,t),
    \label{eq:tdse}
\end{align}
where $u_{b,k}(x,t)$ are periodic part of the Bloch orbitals with the band index $b$ and the Bloch wavenumber
$k$. Here, $A(t)$ is the spatially-uniform time-dependent vector potential and $v(x)$ 
is the single-particle potential. For the single-particle potential $v(x)$,
we employ the Mathieu-type lattice potential
\begin{align}
    V(x) = V_0\cos \left ( \frac{2\pi}{L}x \right ) \label{e:crystal_potential},
\end{align}
where the potential height $V_0$ is set to $0.37$~a.u.
and the lattice constant $L$ is set to $8$~a.u.

In this work, the system is discretized by grid points
for both real and crystal momentum spaces. The real-space unit-cell is discretized with
$30$ equally spaced grid points,
and the first Brillouin zone is discretized with $352$ uniformly spaced grid points.
For the time propagation of the time-dependent Schr\"odigner equation we employ the Taylor expansion scheme \cite{PhysRevB.54.4484}
with the single time step $\Delta t$ set to $1$~attosecond.

\begin{figure}[htbp]
\includegraphics[width=\columnwidth]{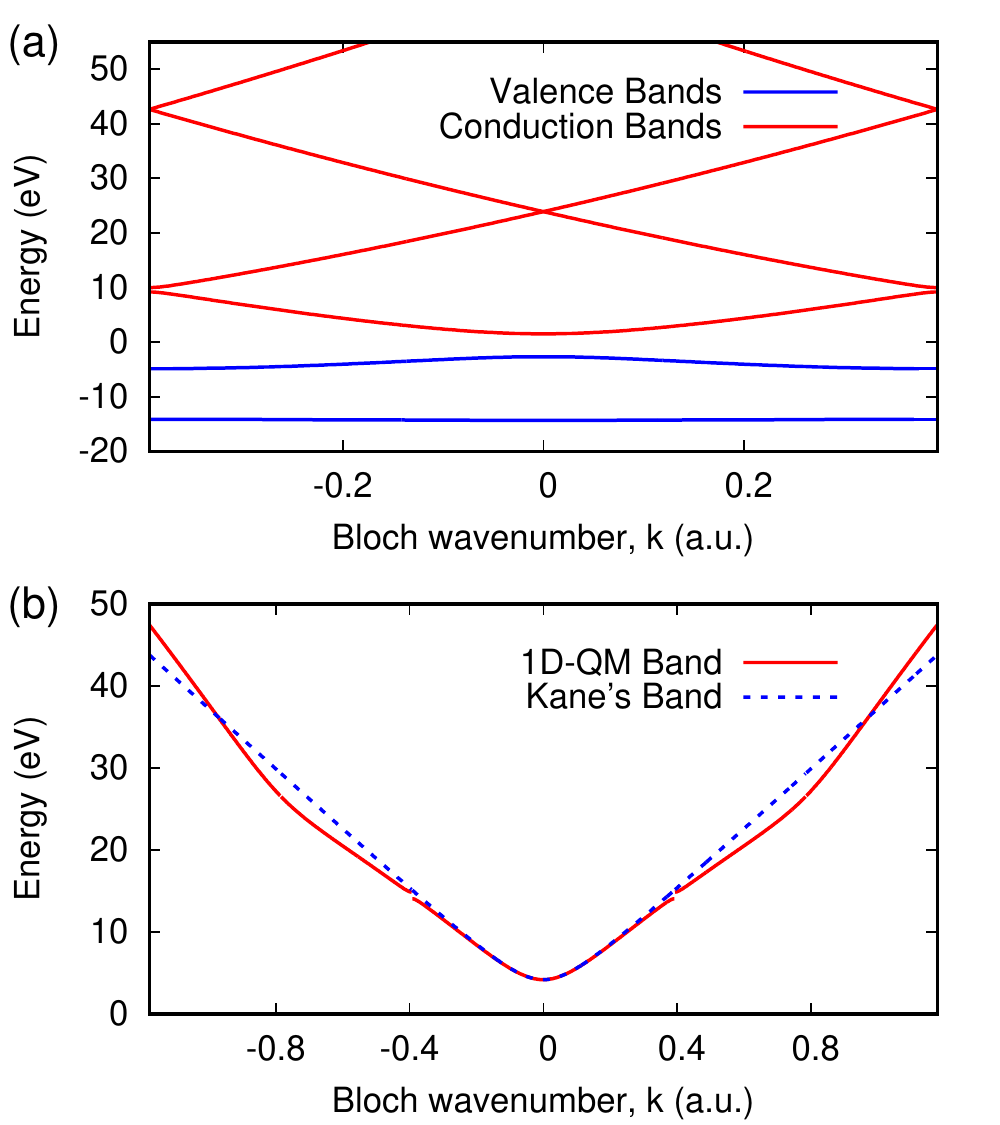}
\caption{\label{fig:BandAndKaneBand}
(a) Electronic band structure of the one-dimensional quantum systems described by
the Matthieu-type lattice potential, Eq.~(\ref{e:crystal_potential}).
The valence bands are shown as the blue lines while the conduction bands are shown
as the red lines.
(b) Comparison of the Kane band (blue-dotted) and the electron-hole
band (red-solid), which is defined as the energy difference
between the conduction bands and the top valence band,
$\epsilon_{c,k}-\epsilon_{v,k}$.
}
\end{figure}

Figure~\ref{fig:BandAndKaneBand}~(a) shows the computed band structure of the 1D-QM model.
The bottom two bands are treated as valence bands, and all the other bands are treated as
the conduction bands. For the comparison with the semi-classical trajectory model,
we fit the band structure of the 1D-QM model by the Kane band model.
Figure~\ref{fig:BandAndKaneBand}~(b) shows the Kane band ${\epsilon_{Kane}(k)}$
and the particle-hole energy band structure of the 1D-QM model.
The particle-hole energy bands are defined by the difference between
the conduction bands and the top valence band as
$\epsilon_{c,k}-\epsilon_{v,k}$, and they are unfolded in
Fig.~\ref{fig:BandAndKaneBand}~(b).
Here, the effective mass $\mu$ of the Kane band model is set to $0.083 m_e$,
and the band gap $\epsilon_g$ is set to $4.18$~eV.
These values are extracted from the 1D-QM model.

We evaluate the HHG spectrum of the 1D-QM model by exactly solving numerically
the time-dependent Schr\"odinger equation, Eq.~(\ref{eq:tdse}), using the following form for the applied vector potential:
\begin{align}
    A(t) = \frac{F_0}{\omega_0}\sin^4 \left (\frac{\pi}{\tau}t \right ) \sin(\omega_0 t)
    \label{eq:laser_field}
\end{align}
in the duration $0<t<\tau$ and zero outside.
Here, $\tau$ is the full duration of the pulse.
In this work, we set the mean photon energy $\hbar \omega_0$ to $387$~meV (with corresponding wavelength $3200$~nm), and the full duration of the pulse $\tau$ to $96.1$~fs (equivalent to $9$ periods of the mean photon frequency) according to
the previous work \cite{Ikemachi_2017_Trajectory}.

During the time propagation, the electric current $J(t)$ can be evaluated by
\begin{align}
    J(t) &= -\sum^2_{b=1} \frac{1}{2\pi}\int^{\pi/L}_{-\pi/L}  dk \nonumber \\
&\times \int^L_0 dx
u^*_{bk}(x,t) \left [ -i \frac{\partial}{\partial x} + k + A(t) \right ]
u_{b,k}(x,t).
    \label{eq:current}
\end{align}
Furthermore, the HHG spectrum $I(\omega)$ can be evaluated as the Fourier transform
of the current as
\begin{align}
I(\omega) \sim \omega^2 \left|
\int^{T_{pulse}}_0 dt \sin^4 \left (\frac{\pi}{\tau} t \right )J(t) e^{i\omega t}
\right|^2,
\end{align}
where the Fourier transform of the current is evaluated with the same envelope function
as the applied laser pulse.

\begin{figure}[htbp]
\includegraphics[width=\columnwidth]{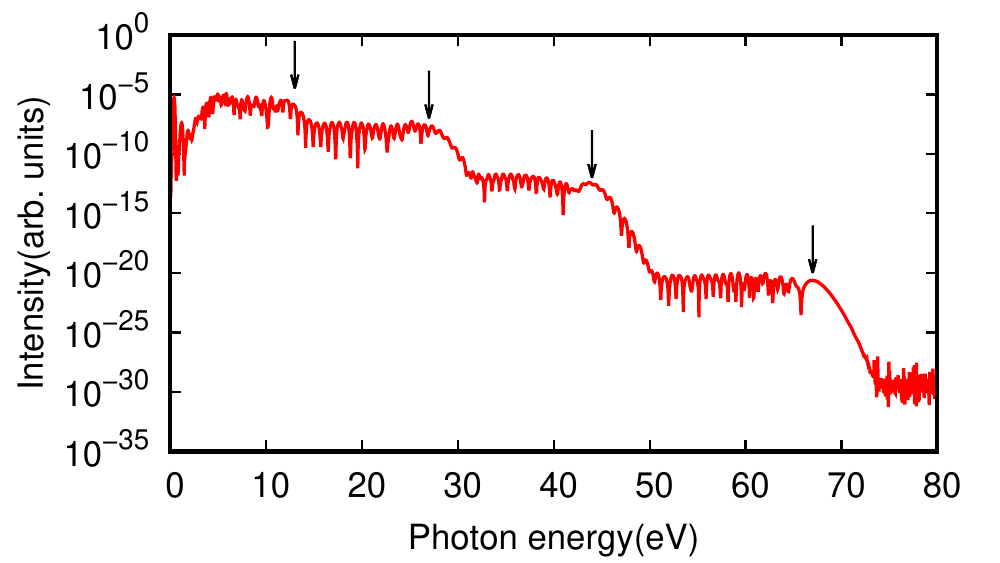}
\caption{\label{fig:hhg_spectrum}
HHG spectrum computed by the quantum simulation with the field strength
of $F_0 = 0.165$ V/\text{\AA}.
The cutoff energy of multiple plateaus are
indicated by black arrows around $13$, $28$, $44$, and $68$~eV.}
\end{figure}

Figure~\ref{fig:hhg_spectrum} shows the computed power spectrum of the HHG from
1D-QM model with the field strength of $F_0 = 0.165$ V/\text{\AA},
showing the clear multiple plateaus.
The multi-plateau feature was investigated with the $k$-space semi-classical trajectory
model, and it was explained by the ladder climbing process in the band structure
\cite{Ikemachi_2017_Trajectory}.
In this subsection, we explore the multi-plateau feature based on the real-space
semi-classical trajectory with the scattering effect.

\begin{figure}[htbp]
\includegraphics[width=\columnwidth]{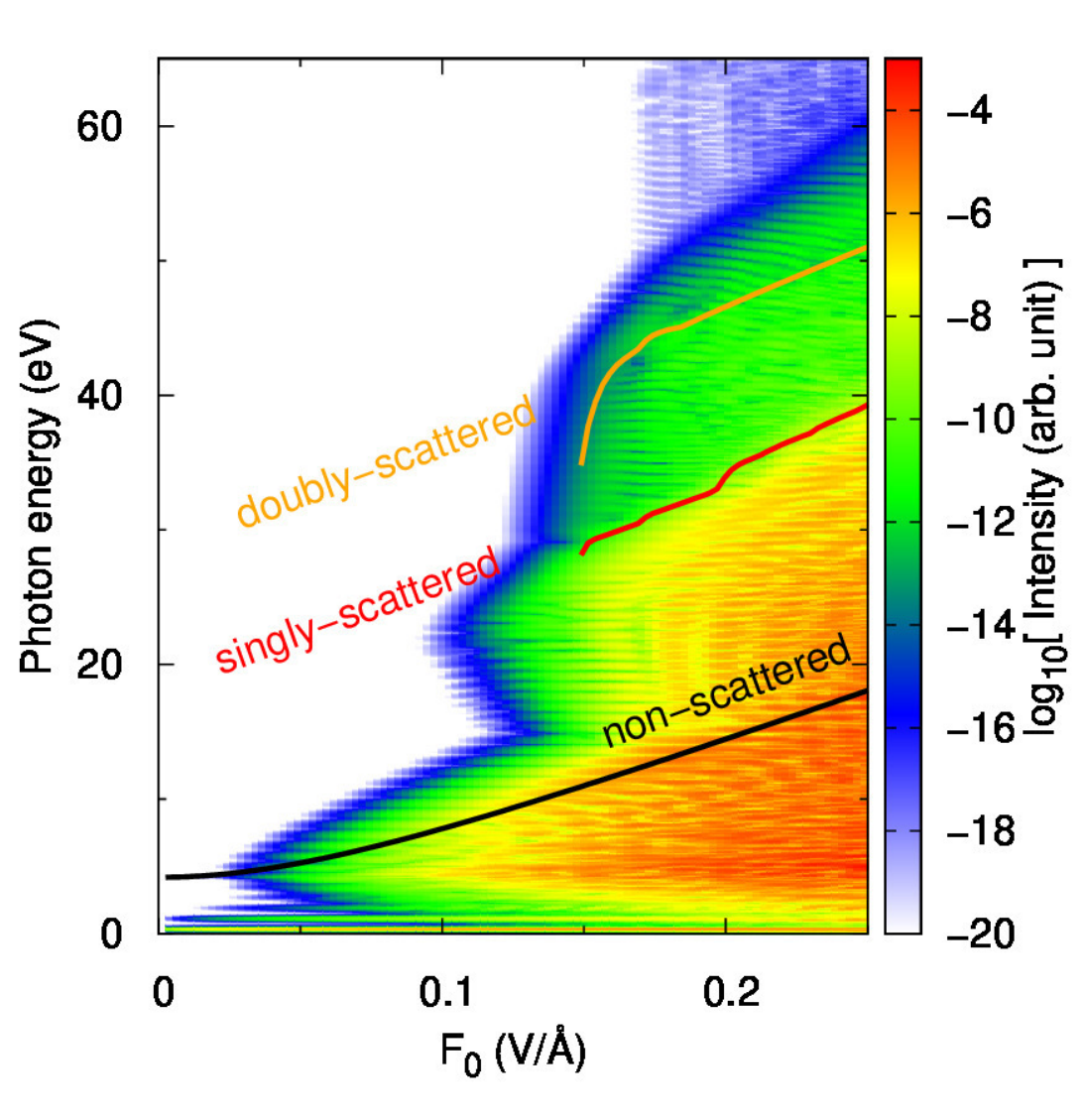}
\caption{\label{fig:hhg_vs_field}
Spectra of HHG computed by the 1D-QM simulation as functions of the field strength, $F_0$.
The cutoff energies computed with the semi-classical trajectory model are also shown:
The black line shows the result without scattering, the red line shows
that of singly-scattered trajectories, and the orange line shows that
of doubly-scattered trajectories.
}
\end{figure}

In our model we explore the multi-plateau feature based on both the $k$-space and the real-space
semi-classical trajectory with the scattering effect.
Figure~\ref{fig:hhg_vs_field} shows the HHG spectra as functions of the applied field
strength $F_0$. One can see the formation of the multi-plateau feature with increase
of the field strength.
To assess the semi-classical trajectory model, the computed cutoff energy
with the Kane band model is also shown as the black-solid line (non-scattered),
which is nothing but the red-solid line shown in Fig.~\ref{fig:cutoff}.
One sees that the first cutoff of the 1D-QM model is captured by
the semi-classical trajectory model without scattering (black line).

To study the role of scattering, we evaluate the maximum recombination energy among
all possible scattered trajectories as the cutoff energy of the scattered trajectory.
In Fig.~\ref{fig:hhg_vs_field}, the red line shows the maximum recombination energy
among the single-scattered trajectories, and the orange line shows that among
the double-scattered trajectories.
By comparing the cutoff energies of the scattered trajectories with
the HHG spectra of the 1D-QM model, one sees that the single-scattering trajectories
(red line) provide the second cutoff of the 1D-QM simulation
while the double-scattering trajectories provide the third cutoff.
Therefore, the formation of the multi-plateau feature in the HHG spectrum
can be understood as the consequence of Umklapp scattering.
This real-space scattering interpretation is a complementary picture of
the previous ladder climbing picture in the $k$-space \cite{Ikemachi_2017_Trajectory}.

\begin{figure}[htbp]
\includegraphics[width = \columnwidth]{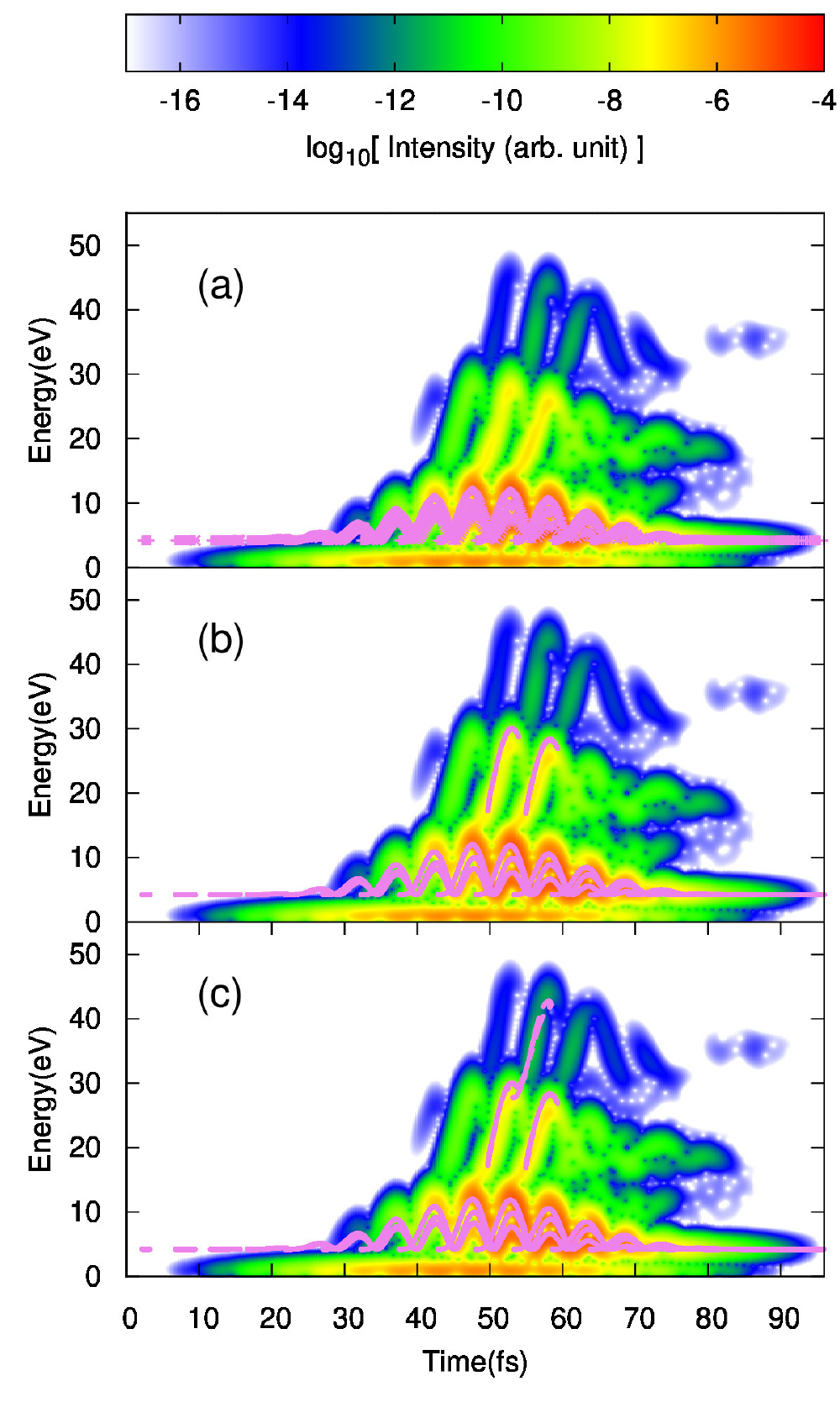}
\caption{\label{fig:TFA_Scatter}
Temporal evolution of the HHG computed with the 1D-QM simulation.
The purple dots describe the emitted photon energy and the emission timing,
computed by the semi-classical trajectory model.
Each panel shows the contribution from semi-classical trajectories with
different number of scattering: (a)~No scattering,(b)~single scattering, and
(c)~double scattering.
}
\end{figure}

To study further details of the scattering effect in the HHG from solids,
we elucidate the temporal structure of HHG. For this purpose, we perform the Gabor
transformation to the current $J(t)$ with the time window function
whose the full width of the half maximum (FWHM) is $1.78$~fs.
The computed temporal evolution of HHG from the 1D-QM model is
shown in Fig.~\ref{fig:TFA_Scatter}. Here, the field strength $F_0$ is set to $0.165$ V/\AA.
Note that, in all the panels of Fig.~\ref{fig:TFA_Scatter}, the same result of
the 1D-QM model is shown. In addition to the 1D-QM result, the recombination
energy and timing evaluated by the semi-classical trajectory model of different number of scattering
are depicted in different panels; The panel~(a), (b), and (c) show the results of non-scattered, singly-scattered, and doubly-scattered trajectories, respectively.

As seen from Fig.~\ref{fig:TFA_Scatter}~(a), the non-scattered trajectories
contribute only to the first plateau of HHG. In contrast,
in Fig.~\ref{fig:TFA_Scatter}~(b), the single-scattered trajectories
show the contribution to the second plateau (around $13$ to $28$ eV). Furthermore, from Fig.~\ref{fig:TFA_Scatter}~(c), the double-scattered trajectories
contribute to the formation of the third plateau (around $28$ to $44$ eV) .
Therefore, Umklapp scattering processes open higher energy channels
for the trajectory dynamics, resulting in the multi-plateau feature as a consequence
of the multiple scatterings.

In Fig.~\ref{fig:TFA_Scatter}~(a) to (c), one sees that the real-space trajectory model
fairly captures the recombination energy and timings.
This is a distinct feature from the $k$-space trajectory model where the recombination
timing is arbitrary and can occur any instance of time.
This fact indicates the importance of the real-space trajectory picture 
to describe the HHG even with scattering process.

Note that the present real-space trajectory model does not capture all the features
of the 1D-QM simulation. For example, in Fig.\ref{fig:TFA_Scatter}~(c) semi-classical model
fails to reproduce the signals on the left side of the classical prediction
in the second plateau or the signal on the right side of the classical prediction
in the third plateau.
These additional features can be understood by
the quantum wavepacket effect. In the present semi-classical model, the excitation
occurs only at the optical gap, and the scattering takes place only at the band crossing points.
However, in the quantum system, the excitation and scattering can occur with finite
width in the $k$-space. As a result, the full quantum systems can involve
more trajectories 
and add the additional features to the semi-classical model.

\subsection{Loss of trajectories by electron scattering in the HHG
\label{subsec:mean_free_path}}

In the above analysis, we investigated the role of Umklapp scattering in one-dimensional
systems. Because only the forward or backward scatterings are allowed in the one-dimensional
systems, the scattered trajectory can be recombined with relatively high probability.
In contrast, the contribution from scattered trajectories in two- and three-dimensional
systems is expected to be significantly suppressed since trajectories can be scattered
into a variety of directions, thus reducing the probability of recombination for returning
trajectories.
In addition to the consideration on dimensionality, there are many scattering processes in solids other than Umklapp scattering.
These scattering processes may also play an important role in solid-phase HHG.

In order to assess the impact of the suppression of HHG by various scattering processes
in the higher-dimensional space,
we consider a simple mean-free-path model instead of the trajectory branching mentioned above.
In the mean-free-path model, we compute the trajectory length $l_{d}(t)$ as
\begin{align}
l_{d}(t) = \int^t_{t_0}dt' \left | \vecb v\left (\vecb K(t') \right )\right |.
\end{align}
Furthermore, we simply assume that semi-classical trajectories do not contribute to
HHG anymore once the trajectory length reaches a given mean-free-path length, $l_{MFP}$.
Integrating this destructive contribution of scattering into the semi-classical trajectory
model instead of the trajectory-branching process, we evaluate the maximum recombination
energy among all trajectories before their trajectory length reaches the mean-free-path length.

\begin{figure}[htbp]
\includegraphics[width=\columnwidth]{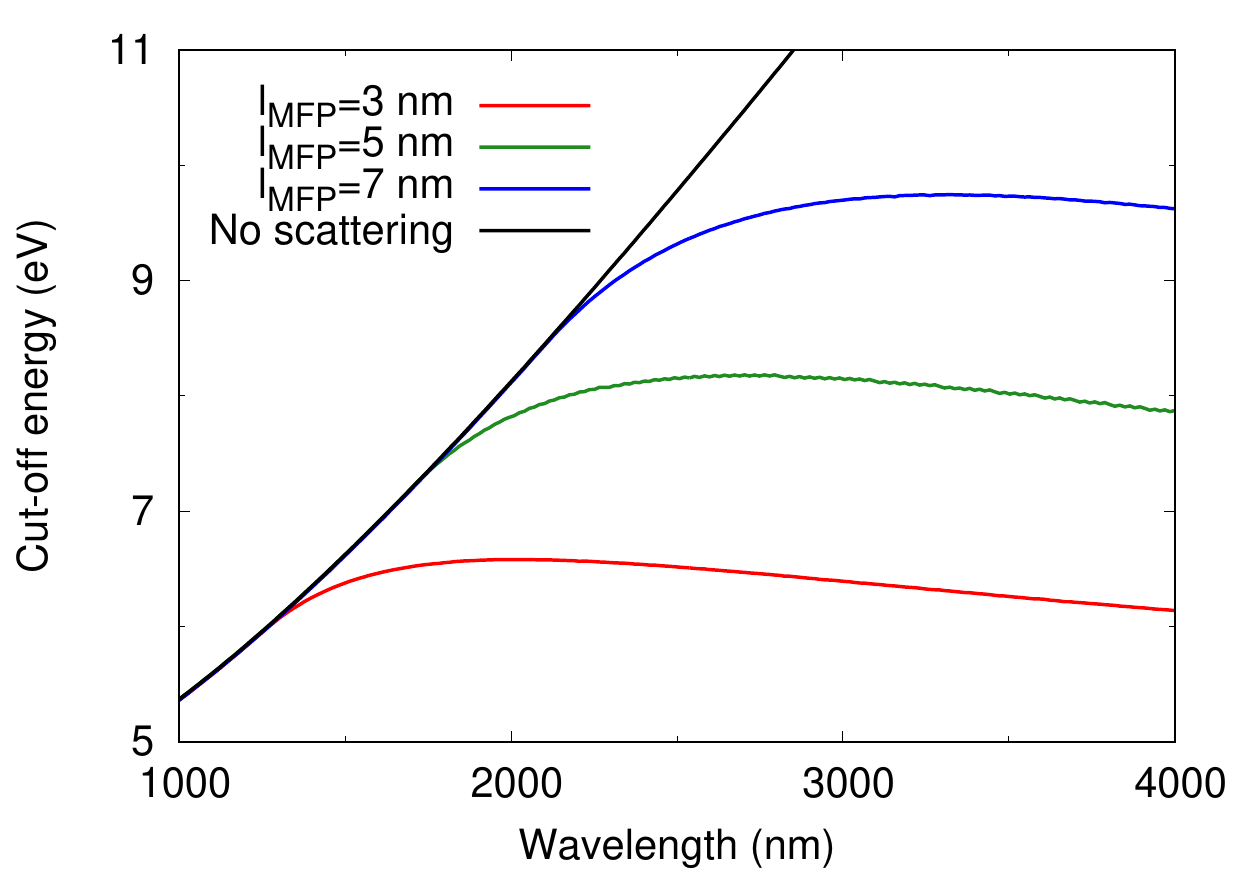}
\caption{\label{fig:CutOff_TSM_MFP}
Wavelength dependence of the cutoff energy of HHG, evaluated by the semi-classical
trajectory model with the mean-free-path approximation.
The results with several mean-free-path length $l_{MFP}$ are described.
The result without scattering process is also shown as the black line.
}
\end{figure}

Figure~\ref{fig:CutOff_TSM_MFP} shows the computed cutoff energy from
the semi-classical trajectory model for different mean-free-path length $l_{MFP}$ as functions
of the applied laser wavelength under the fixed field strength $F_0 = 0.165$ V/\AA. As seen from Fig.~\ref{fig:CutOff_TSM_MFP}, the trajectory model without scattering (black)
shows almost linear dependency in the long wavelength region.
This behavior can be understood as the asymptotic linear dispersion of the Kane band model
in the large Bloch wavevector region, as described in Eq.~(\ref{eq:cutoff_kane}).

Once the loss of trajectories is introduced via the mean-free-path approximation,
the semi-classical trajectory model shows the saturation of the HHG
cutoff energy in the long wavelength region.
This saturation can be understood by a fact that,
even if laser fields with longer wavelength
can induce higher energy trajectories, such higher-energy trajectories have
longer travel distance and they are lost by the scattering process.

One sees that the cutoff energy of the trajectory model with the scattering effect
in Fig.~\ref{fig:CutOff_TSM_MFP} becomes almost constant in the long wavelength region.
This behavior is nothing but the wavelength independence of the HHG cutoff energy,
and it is consistent with one of the features of the HHG from solids
\cite{Ghimire2011,Luu2015,Nicolas_2017_Impact}.
Thus, the loss of trajectories by scattering can be one of the physical mechanisms
of the wavelength independence of the cutoff energy of HHG from solids.
In order to clarify the details of the trajectory loss effect, the comparison
of the trajectory model with the three-dimensional realistic simulations 
such as the \textit{ab-initio} time-dependent density functional theory simulation,
which also captures the wavelength independence of the HHG cutoff energy
\cite{Nicolas_2017_Impact}, will be useful.
However, such analysis is beyond the scope of the present work, and will be
investigated in the future.

\section{Summary \label{sec:summary}}

We studied the effect of electron scattering in the HHG from solids
based on the semi-classical trajectory description.
We first extended the solid-state semi-classical trajectory model
\cite{Vampa_2015_semiclassical} by integrating the Umklapp scattering with
the Kane band model \cite{Kane_1957_Band}. The extended model has been examined by comparing with
the one-dimensional quantum dynamical simulation used in
Ref.~\cite{Ikemachi_2017_Trajectory}.
As a result, the multi-plateau feature of the HHG spectra of the one-dimensional quantum
model has been fairly captured by the contribution from the multiple
Umklapp-scattered trajectories under the laser field acceleration.
Therefore, we concluded that the multi-plateau feature is the consequence of
the Umklapp scattering. In the previous work based on the $k$-space trajectory model
\cite{Ikemachi_2017_Trajectory}, the same multi-plateau feature has been interpreted
as the consequence of the ladder-climbing process in the electronic band structure.
The two interpretations based on the real-space and $k$-space trajectories
are equivalent but offers different views into the same phenomenon: the ladder-climbing
process in $k$-space can be seen as the dynamics with Umklapp scattering in real-space.

We further examine the effect of Umklapp scattering by evaluating the recombination timing of electron-hole pairs and the emitted photon energy.
By comparing the results of the semi-classical trajectory model with the temporal
structure of the HHG spectra of the one-dimensional quantum model, we confirm that
the semi-classical trajectory model with Umklapp scattering properly captures
the timing and energy of the HHG from solids. This fact demonstrates that 
the semi-classical real-space trajectories play an important role in the microscopic
mechanism of the HHG from solids because the recombination timing is not determined in
the purely $k$-space trajectory model, where the recombination can occur at any time
instance.

Then, we explored other consequence of the electron scattering, considering that
scattered trajectories in a higher-dimensional space 
are less likely to recombine. To take into account this effect,
we evaluate the travel distance of each trajectory and disregard
trajectories if their travel distance reaches a given mean-free-path length.
We evaluated the wavelength dependence of the HHG cutoff energy
by the semi-classical trajectory model with the mean-free-path approximation.
As a result, we found that the cutoff energy is significantly suppressed
under the longer wavelength laser driving since the longer wavelength laser field
tends to induce longer travel distance for trajectories.
In addition, we found that the cutoff energy is almost independent of
the laser wavelength once the wavelength becomes long enough.
This wavelength independence is consistent with a feature of the reported HHG
from solids \cite{Ghimire2011,Luu2015,Nicolas_2017_Impact}.

The above findings indicate that the real-space trajectory combined
with scattering processes plays an essential role in the HHG from solids.
Thus, the optical control of the real-space trajectories under scattering processes
may further open a way to enhance and control the HHG from solids.

\begin{acknowledgments}
This work was supported by the European Research Council (ERC-2015-AdG694097),
the Cluster of Excellence 'Advanced Imaging of Matter' (AIM),
and JST-CREST under Grant No. JP-MJCR16N5.
S.A.S. gratefully acknowledges the fellowship from the Alexander von Humboldt Foundation.
\end{acknowledgments}

\bibliography{ref}

\begin{thebibliography}{44}%
\makeatletter
\providecommand \@ifxundefined [1]{%
 \@ifx{#1\undefined}
}%
\providecommand \@ifnum [1]{%
 \ifnum #1\expandafter \@firstoftwo
 \else \expandafter \@secondoftwo
 \fi
}%
\providecommand \@ifx [1]{%
 \ifx #1\expandafter \@firstoftwo
 \else \expandafter \@secondoftwo
 \fi
}%
\providecommand \natexlab [1]{#1}%
\providecommand \enquote  [1]{``#1''}%
\providecommand \bibnamefont  [1]{#1}%
\providecommand \bibfnamefont [1]{#1}%
\providecommand \citenamefont [1]{#1}%
\providecommand \href@noop [0]{\@secondoftwo}%
\providecommand \href [0]{\begingroup \@sanitize@url \@href}%
\providecommand \@href[1]{\@@startlink{#1}\@@href}%
\providecommand \@@href[1]{\endgroup#1\@@endlink}%
\providecommand \@sanitize@url [0]{\catcode `\\12\catcode `\$12\catcode
  `\&12\catcode `\#12\catcode `\^12\catcode `\_12\catcode `\%12\relax}%
\providecommand \@@startlink[1]{}%
\providecommand \@@endlink[0]{}%
\providecommand \url  [0]{\begingroup\@sanitize@url \@url }%
\providecommand \@url [1]{\endgroup\@href {#1}{\urlprefix }}%
\providecommand \urlprefix  [0]{URL }%
\providecommand \Eprint [0]{\href }%
\providecommand \doibase [0]{http://dx.doi.org/}%
\providecommand \selectlanguage [0]{\@gobble}%
\providecommand \bibinfo  [0]{\@secondoftwo}%
\providecommand \bibfield  [0]{\@secondoftwo}%
\providecommand \translation [1]{[#1]}%
\providecommand \BibitemOpen [0]{}%
\providecommand \bibitemStop [0]{}%
\providecommand \bibitemNoStop [0]{.\EOS\space}%
\providecommand \EOS [0]{\spacefactor3000\relax}%
\providecommand \BibitemShut  [1]{\csname bibitem#1\endcsname}%
\let\auto@bib@innerbib\@empty
\bibitem [{\citenamefont {Mourou}\ \emph {et~al.}(2012)\citenamefont {Mourou},
  \citenamefont {Fisch}, \citenamefont {Malkin}, \citenamefont {Toroker},
  \citenamefont {Khazanov}, \citenamefont {Sergeev}, \citenamefont {Tajima},\
  and\ \citenamefont {Garrec}}]{MOUROU2012720}%
  \BibitemOpen
  \bibfield  {author} {\bibinfo {author} {\bibfnamefont {G.}~\bibnamefont
  {Mourou}}, \bibinfo {author} {\bibfnamefont {N.}~\bibnamefont {Fisch}},
  \bibinfo {author} {\bibfnamefont {V.}~\bibnamefont {Malkin}}, \bibinfo
  {author} {\bibfnamefont {Z.}~\bibnamefont {Toroker}}, \bibinfo {author}
  {\bibfnamefont {E.}~\bibnamefont {Khazanov}}, \bibinfo {author}
  {\bibfnamefont {A.}~\bibnamefont {Sergeev}}, \bibinfo {author} {\bibfnamefont
  {T.}~\bibnamefont {Tajima}}, \ and\ \bibinfo {author} {\bibfnamefont {B.~L.}\
  \bibnamefont {Garrec}},\ }\href {\doibase
  https://doi.org/10.1016/j.optcom.2011.10.089} {\bibfield  {journal} {\bibinfo
   {journal} {Optics Communications}\ }\textbf {\bibinfo {volume} {285}},\
  \bibinfo {pages} {720 } (\bibinfo {year} {2012})}\BibitemShut {NoStop}%
\bibitem [{\citenamefont {Krausz}\ and\ \citenamefont
  {Stockman}(2014)}]{Krausz_2014_Attosecond}%
  \BibitemOpen
  \bibfield  {author} {\bibinfo {author} {\bibfnamefont {F.}~\bibnamefont
  {Krausz}}\ and\ \bibinfo {author} {\bibfnamefont {M.~I.}\ \bibnamefont
  {Stockman}},\ }\href {\doibase 10.1038/nphoton.2014.28} {\bibfield  {journal}
  {\bibinfo  {journal} {Nature Photonics}\ }\textbf {\bibinfo {volume} {8}},\
  \bibinfo {pages} {205} (\bibinfo {year} {2014})}\BibitemShut {NoStop}%
\bibitem [{\citenamefont {Basov}\ \emph {et~al.}(2017)\citenamefont {Basov},
  \citenamefont {Averitt},\ and\ \citenamefont {Hsieh}}]{Basov2017}%
  \BibitemOpen
  \bibfield  {author} {\bibinfo {author} {\bibfnamefont {D.~N.}\ \bibnamefont
  {Basov}}, \bibinfo {author} {\bibfnamefont {R.~D.}\ \bibnamefont {Averitt}},
  \ and\ \bibinfo {author} {\bibfnamefont {D.}~\bibnamefont {Hsieh}},\ }\href
  {\doibase 10.1038/nmat5017} {\bibfield  {journal} {\bibinfo  {journal}
  {Nature Materials}\ }\textbf {\bibinfo {volume} {16}},\ \bibinfo {pages}
  {1077} (\bibinfo {year} {2017})}\BibitemShut {NoStop}%
\bibitem [{\citenamefont {Flick}\ \emph {et~al.}(2017)\citenamefont {Flick},
  \citenamefont {Ruggenthaler}, \citenamefont {Appel},\ and\ \citenamefont
  {Rubio}}]{Flick3026}%
  \BibitemOpen
  \bibfield  {author} {\bibinfo {author} {\bibfnamefont {J.}~\bibnamefont
  {Flick}}, \bibinfo {author} {\bibfnamefont {M.}~\bibnamefont {Ruggenthaler}},
  \bibinfo {author} {\bibfnamefont {H.}~\bibnamefont {Appel}}, \ and\ \bibinfo
  {author} {\bibfnamefont {A.}~\bibnamefont {Rubio}},\ }\href {\doibase
  10.1073/pnas.1615509114} {\bibfield  {journal} {\bibinfo  {journal}
  {Proceedings of the National Academy of Sciences}\ }\textbf {\bibinfo
  {volume} {114}},\ \bibinfo {pages} {3026} (\bibinfo {year}
  {2017})}\BibitemShut {NoStop}%
\bibitem [{\citenamefont {Butcher}\ and\ \citenamefont
  {Cotter}(1990)}]{butcher1990elements}%
  \BibitemOpen
  \bibfield  {author} {\bibinfo {author} {\bibfnamefont {P.~N.}\ \bibnamefont
  {Butcher}}\ and\ \bibinfo {author} {\bibfnamefont {D.}~\bibnamefont
  {Cotter}},\ }\href@noop {} {\emph {\bibinfo {title} {The elements of
  nonlinear optics}}},\ Vol.~\bibinfo {volume} {9}\ (\bibinfo  {publisher}
  {Cambridge university press},\ \bibinfo {year} {1990})\BibitemShut {NoStop}%
\bibitem [{\citenamefont {Boyd}(2008)}]{2008v}%
  \BibitemOpen
  \bibfield  {author} {\bibinfo {author} {\bibfnamefont {R.~W.}\ \bibnamefont
  {Boyd}},\ }\href {\doibase
  https://doi.org/10.1016/B978-0-12-369470-6.00020-4} {\emph {\bibinfo {title}
  {Nonlinear Optics}}}\ (\bibinfo  {publisher} {Academic Press, New York},\
  \bibinfo {year} {2008})\BibitemShut {NoStop}%
\bibitem [{\citenamefont {Hentschel}\ \emph {et~al.}(2001)\citenamefont
  {Hentschel}, \citenamefont {Kienberger}, \citenamefont {Spielmann},
  \citenamefont {Reider}, \citenamefont {Milosevic}, \citenamefont {Brabec},
  \citenamefont {Corkum}, \citenamefont {Heinzmann}, \citenamefont {Drescher},\
  and\ \citenamefont {Krausz}}]{Hentschel_2001_Attosecond}%
  \BibitemOpen
  \bibfield  {author} {\bibinfo {author} {\bibfnamefont {M.}~\bibnamefont
  {Hentschel}}, \bibinfo {author} {\bibfnamefont {R.}~\bibnamefont
  {Kienberger}}, \bibinfo {author} {\bibfnamefont {C.}~\bibnamefont
  {Spielmann}}, \bibinfo {author} {\bibfnamefont {G.~A.}\ \bibnamefont
  {Reider}}, \bibinfo {author} {\bibfnamefont {N.}~\bibnamefont {Milosevic}},
  \bibinfo {author} {\bibfnamefont {T.}~\bibnamefont {Brabec}}, \bibinfo
  {author} {\bibfnamefont {P.}~\bibnamefont {Corkum}}, \bibinfo {author}
  {\bibfnamefont {U.}~\bibnamefont {Heinzmann}}, \bibinfo {author}
  {\bibfnamefont {M.}~\bibnamefont {Drescher}}, \ and\ \bibinfo {author}
  {\bibfnamefont {F.}~\bibnamefont {Krausz}},\ }\href {\doibase
  10.1038/35107000} {\bibfield  {journal} {\bibinfo  {journal} {Nature}\
  }\textbf {\bibinfo {volume} {414}},\ \bibinfo {pages} {509} (\bibinfo {year}
  {2001})}\BibitemShut {NoStop}%
\bibitem [{\citenamefont {Pfeifer}\ \emph {et~al.}(2006)\citenamefont
  {Pfeifer}, \citenamefont {Spielmann},\ and\ \citenamefont
  {Gerber}}]{Pfeifer_2006_Femtosecond}%
  \BibitemOpen
  \bibfield  {author} {\bibinfo {author} {\bibfnamefont {T.}~\bibnamefont
  {Pfeifer}}, \bibinfo {author} {\bibfnamefont {C.}~\bibnamefont {Spielmann}},
  \ and\ \bibinfo {author} {\bibfnamefont {G.}~\bibnamefont {Gerber}},\ }\href
  {\doibase 10.1088/0034-4885/69/2/r04} {\bibfield  {journal} {\bibinfo
  {journal} {Reports on Progress in Physics}\ }\textbf {\bibinfo {volume}
  {69}},\ \bibinfo {pages} {443} (\bibinfo {year} {2006})}\BibitemShut
  {NoStop}%
\bibitem [{\citenamefont {Krausz}\ and\ \citenamefont
  {Ivanov}(2009)}]{Krausz_2009_Attosecond}%
  \BibitemOpen
  \bibfield  {author} {\bibinfo {author} {\bibfnamefont {F.}~\bibnamefont
  {Krausz}}\ and\ \bibinfo {author} {\bibfnamefont {M.}~\bibnamefont
  {Ivanov}},\ }\href {\doibase 10.1103/RevModPhys.81.163} {\bibfield  {journal}
  {\bibinfo  {journal} {Rev. Mod. Phys.}\ }\textbf {\bibinfo {volume} {81}},\
  \bibinfo {pages} {163} (\bibinfo {year} {2009})}\BibitemShut {NoStop}%
\bibitem [{\citenamefont {Franken}\ \emph {et~al.}(1961)\citenamefont
  {Franken}, \citenamefont {Hill}, \citenamefont {Peters},\ and\ \citenamefont
  {Weinreich}}]{PhysRevLett.7.118}%
  \BibitemOpen
  \bibfield  {author} {\bibinfo {author} {\bibfnamefont {P.~A.}\ \bibnamefont
  {Franken}}, \bibinfo {author} {\bibfnamefont {A.~E.}\ \bibnamefont {Hill}},
  \bibinfo {author} {\bibfnamefont {C.~W.}\ \bibnamefont {Peters}}, \ and\
  \bibinfo {author} {\bibfnamefont {G.}~\bibnamefont {Weinreich}},\ }\href
  {\doibase 10.1103/PhysRevLett.7.118} {\bibfield  {journal} {\bibinfo
  {journal} {Phys. Rev. Lett.}\ }\textbf {\bibinfo {volume} {7}},\ \bibinfo
  {pages} {118} (\bibinfo {year} {1961})}\BibitemShut {NoStop}%
\bibitem [{\citenamefont {Krause}\ \emph {et~al.}(1992)\citenamefont {Krause},
  \citenamefont {Schafer},\ and\ \citenamefont {Kulander}}]{Krause_1992_High}%
  \BibitemOpen
  \bibfield  {author} {\bibinfo {author} {\bibfnamefont {J.~L.}\ \bibnamefont
  {Krause}}, \bibinfo {author} {\bibfnamefont {K.~J.}\ \bibnamefont {Schafer}},
  \ and\ \bibinfo {author} {\bibfnamefont {K.~C.}\ \bibnamefont {Kulander}},\
  }\href {\doibase 10.1103/PhysRevLett.68.3535} {\bibfield  {journal} {\bibinfo
   {journal} {Phys. Rev. Lett.}\ }\textbf {\bibinfo {volume} {68}},\ \bibinfo
  {pages} {3535} (\bibinfo {year} {1992})}\BibitemShut {NoStop}%
\bibitem [{\citenamefont {Schafer}\ \emph {et~al.}(1993)\citenamefont
  {Schafer}, \citenamefont {Yang}, \citenamefont {DiMauro},\ and\ \citenamefont
  {Kulander}}]{Schafer_1993_Above}%
  \BibitemOpen
  \bibfield  {author} {\bibinfo {author} {\bibfnamefont {K.~J.}\ \bibnamefont
  {Schafer}}, \bibinfo {author} {\bibfnamefont {B.}~\bibnamefont {Yang}},
  \bibinfo {author} {\bibfnamefont {L.~F.}\ \bibnamefont {DiMauro}}, \ and\
  \bibinfo {author} {\bibfnamefont {K.~C.}\ \bibnamefont {Kulander}},\ }\href
  {\doibase 10.1103/PhysRevLett.70.1599} {\bibfield  {journal} {\bibinfo
  {journal} {Phys. Rev. Lett.}\ }\textbf {\bibinfo {volume} {70}},\ \bibinfo
  {pages} {1599} (\bibinfo {year} {1993})}\BibitemShut {NoStop}%
\bibitem [{\citenamefont {Lewenstein}\ \emph
  {et~al.}(1994{\natexlab{a}})\citenamefont {Lewenstein}, \citenamefont
  {Balcou}, \citenamefont {Ivanov}, \citenamefont {L’Huillier},\ and\
  \citenamefont {Corkum}}]{Lewenstein_1994_Theory}%
  \BibitemOpen
  \bibfield  {author} {\bibinfo {author} {\bibfnamefont {M.}~\bibnamefont
  {Lewenstein}}, \bibinfo {author} {\bibfnamefont {P.}~\bibnamefont {Balcou}},
  \bibinfo {author} {\bibfnamefont {M.~Y.}\ \bibnamefont {Ivanov}}, \bibinfo
  {author} {\bibfnamefont {A.}~\bibnamefont {L’Huillier}}, \ and\ \bibinfo
  {author} {\bibfnamefont {P.~B.}\ \bibnamefont {Corkum}},\ }\href {\doibase
  10.1103/PhysRevA.49.2117} {\bibfield  {journal} {\bibinfo  {journal}
  {Physical Review A}\ }\textbf {\bibinfo {volume} {49}},\ \bibinfo {pages}
  {2117} (\bibinfo {year} {1994}{\natexlab{a}})}\BibitemShut {NoStop}%
\bibitem [{\citenamefont {McPherson}\ \emph {et~al.}(1987)\citenamefont
  {McPherson}, \citenamefont {Gibson}, \citenamefont {Jara}, \citenamefont
  {Johann}, \citenamefont {Luk}, \citenamefont {McIntyre}, \citenamefont
  {Boyer},\ and\ \citenamefont {Rhodes}}]{McPherson_1987_Studies}%
  \BibitemOpen
  \bibfield  {author} {\bibinfo {author} {\bibfnamefont {A.}~\bibnamefont
  {McPherson}}, \bibinfo {author} {\bibfnamefont {G.}~\bibnamefont {Gibson}},
  \bibinfo {author} {\bibfnamefont {H.}~\bibnamefont {Jara}}, \bibinfo {author}
  {\bibfnamefont {U.}~\bibnamefont {Johann}}, \bibinfo {author} {\bibfnamefont
  {T.~S.}\ \bibnamefont {Luk}}, \bibinfo {author} {\bibfnamefont {I.~A.}\
  \bibnamefont {McIntyre}}, \bibinfo {author} {\bibfnamefont {K.}~\bibnamefont
  {Boyer}}, \ and\ \bibinfo {author} {\bibfnamefont {C.~K.}\ \bibnamefont
  {Rhodes}},\ }\href {\doibase 10.1364/JOSAB.4.000595} {\bibfield  {journal}
  {\bibinfo  {journal} {J. Opt. Soc. Am. B}\ }\textbf {\bibinfo {volume} {4}},\
  \bibinfo {pages} {595} (\bibinfo {year} {1987})}\BibitemShut {NoStop}%
\bibitem [{\citenamefont {Ferray}\ \emph {et~al.}(1988)\citenamefont {Ferray},
  \citenamefont {L{\textquotesingle}Huillier}, \citenamefont {Li},
  \citenamefont {Lompre}, \citenamefont {Mainfray},\ and\ \citenamefont
  {Manus}}]{Ferray_1988_Multiple}%
  \BibitemOpen
  \bibfield  {author} {\bibinfo {author} {\bibfnamefont {M.}~\bibnamefont
  {Ferray}}, \bibinfo {author} {\bibfnamefont {A.}~\bibnamefont
  {L{\textquotesingle}Huillier}}, \bibinfo {author} {\bibfnamefont {X.~F.}\
  \bibnamefont {Li}}, \bibinfo {author} {\bibfnamefont {L.~A.}\ \bibnamefont
  {Lompre}}, \bibinfo {author} {\bibfnamefont {G.}~\bibnamefont {Mainfray}}, \
  and\ \bibinfo {author} {\bibfnamefont {C.}~\bibnamefont {Manus}},\ }\href
  {\doibase 10.1088/0953-4075/21/3/001} {\bibfield  {journal} {\bibinfo
  {journal} {Journal of Physics B: Atomic, Molecular and Optical Physics}\
  }\textbf {\bibinfo {volume} {21}},\ \bibinfo {pages} {L31} (\bibinfo {year}
  {1988})}\BibitemShut {NoStop}%
\bibitem [{\citenamefont {Itatani}\ \emph {et~al.}(2004)\citenamefont
  {Itatani}, \citenamefont {Levesque}, \citenamefont {Zeidler}, \citenamefont
  {Niikura}, \citenamefont {P{\'e}pin}, \citenamefont {Kieffer}, \citenamefont
  {Corkum},\ and\ \citenamefont {Villeneuve}}]{Itatani2004}%
  \BibitemOpen
  \bibfield  {author} {\bibinfo {author} {\bibfnamefont {J.}~\bibnamefont
  {Itatani}}, \bibinfo {author} {\bibfnamefont {J.}~\bibnamefont {Levesque}},
  \bibinfo {author} {\bibfnamefont {D.}~\bibnamefont {Zeidler}}, \bibinfo
  {author} {\bibfnamefont {H.}~\bibnamefont {Niikura}}, \bibinfo {author}
  {\bibfnamefont {H.}~\bibnamefont {P{\'e}pin}}, \bibinfo {author}
  {\bibfnamefont {J.~C.}\ \bibnamefont {Kieffer}}, \bibinfo {author}
  {\bibfnamefont {P.~B.}\ \bibnamefont {Corkum}}, \ and\ \bibinfo {author}
  {\bibfnamefont {D.~M.}\ \bibnamefont {Villeneuve}},\ }\href {\doibase
  10.1038/nature03183} {\bibfield  {journal} {\bibinfo  {journal} {Nature}\
  }\textbf {\bibinfo {volume} {432}},\ \bibinfo {pages} {867} (\bibinfo {year}
  {2004})}\BibitemShut {NoStop}%
\bibitem [{\citenamefont {Goulielmakis}\ \emph {et~al.}(2010)\citenamefont
  {Goulielmakis}, \citenamefont {Loh}, \citenamefont {Wirth}, \citenamefont
  {Santra}, \citenamefont {Rohringer}, \citenamefont {Yakovlev}, \citenamefont
  {Zherebtsov}, \citenamefont {Pfeifer}, \citenamefont {Azzeer}, \citenamefont
  {Kling}, \citenamefont {Leone},\ and\ \citenamefont
  {Krausz}}]{Goulielmakis2010}%
  \BibitemOpen
  \bibfield  {author} {\bibinfo {author} {\bibfnamefont {E.}~\bibnamefont
  {Goulielmakis}}, \bibinfo {author} {\bibfnamefont {Z.-H.}\ \bibnamefont
  {Loh}}, \bibinfo {author} {\bibfnamefont {A.}~\bibnamefont {Wirth}}, \bibinfo
  {author} {\bibfnamefont {R.}~\bibnamefont {Santra}}, \bibinfo {author}
  {\bibfnamefont {N.}~\bibnamefont {Rohringer}}, \bibinfo {author}
  {\bibfnamefont {V.~S.}\ \bibnamefont {Yakovlev}}, \bibinfo {author}
  {\bibfnamefont {S.}~\bibnamefont {Zherebtsov}}, \bibinfo {author}
  {\bibfnamefont {T.}~\bibnamefont {Pfeifer}}, \bibinfo {author} {\bibfnamefont
  {A.~M.}\ \bibnamefont {Azzeer}}, \bibinfo {author} {\bibfnamefont {M.~F.}\
  \bibnamefont {Kling}}, \bibinfo {author} {\bibfnamefont {S.~R.}\ \bibnamefont
  {Leone}}, \ and\ \bibinfo {author} {\bibfnamefont {F.}~\bibnamefont
  {Krausz}},\ }\href {\doibase 10.1038/nature09212} {\bibfield  {journal}
  {\bibinfo  {journal} {Nature}\ }\textbf {\bibinfo {volume} {466}},\ \bibinfo
  {pages} {739} (\bibinfo {year} {2010})}\BibitemShut {NoStop}%
\bibitem [{\citenamefont {Schultze}\ \emph {et~al.}(2014)\citenamefont
  {Schultze}, \citenamefont {Ramasesha}, \citenamefont {Pemmaraju},
  \citenamefont {Sato}, \citenamefont {Whitmore}, \citenamefont {Gandman},
  \citenamefont {Prell}, \citenamefont {Borja}, \citenamefont {Prendergast},
  \citenamefont {Yabana}, \citenamefont {Neumark},\ and\ \citenamefont
  {Leone}}]{Schultze1348}%
  \BibitemOpen
  \bibfield  {author} {\bibinfo {author} {\bibfnamefont {M.}~\bibnamefont
  {Schultze}}, \bibinfo {author} {\bibfnamefont {K.}~\bibnamefont {Ramasesha}},
  \bibinfo {author} {\bibfnamefont {C.~D.}\ \bibnamefont {Pemmaraju}}, \bibinfo
  {author} {\bibfnamefont {S.~A.}\ \bibnamefont {Sato}}, \bibinfo {author}
  {\bibfnamefont {D.}~\bibnamefont {Whitmore}}, \bibinfo {author}
  {\bibfnamefont {A.}~\bibnamefont {Gandman}}, \bibinfo {author} {\bibfnamefont
  {J.~S.}\ \bibnamefont {Prell}}, \bibinfo {author} {\bibfnamefont {L.~J.}\
  \bibnamefont {Borja}}, \bibinfo {author} {\bibfnamefont {D.}~\bibnamefont
  {Prendergast}}, \bibinfo {author} {\bibfnamefont {K.}~\bibnamefont {Yabana}},
  \bibinfo {author} {\bibfnamefont {D.~M.}\ \bibnamefont {Neumark}}, \ and\
  \bibinfo {author} {\bibfnamefont {S.~R.}\ \bibnamefont {Leone}},\ }\href
  {\doibase 10.1126/science.1260311} {\bibfield  {journal} {\bibinfo  {journal}
  {Science}\ }\textbf {\bibinfo {volume} {346}},\ \bibinfo {pages} {1348}
  (\bibinfo {year} {2014})}\BibitemShut {NoStop}%
\bibitem [{\citenamefont {Lucchini}\ \emph {et~al.}(2016)\citenamefont
  {Lucchini}, \citenamefont {Sato}, \citenamefont {Ludwig}, \citenamefont
  {Herrmann}, \citenamefont {Volkov}, \citenamefont {Kasmi}, \citenamefont
  {Shinohara}, \citenamefont {Yabana}, \citenamefont {Gallmann},\ and\
  \citenamefont {Keller}}]{Lucchini916}%
  \BibitemOpen
  \bibfield  {author} {\bibinfo {author} {\bibfnamefont {M.}~\bibnamefont
  {Lucchini}}, \bibinfo {author} {\bibfnamefont {S.~A.}\ \bibnamefont {Sato}},
  \bibinfo {author} {\bibfnamefont {A.}~\bibnamefont {Ludwig}}, \bibinfo
  {author} {\bibfnamefont {J.}~\bibnamefont {Herrmann}}, \bibinfo {author}
  {\bibfnamefont {M.}~\bibnamefont {Volkov}}, \bibinfo {author} {\bibfnamefont
  {L.}~\bibnamefont {Kasmi}}, \bibinfo {author} {\bibfnamefont
  {Y.}~\bibnamefont {Shinohara}}, \bibinfo {author} {\bibfnamefont
  {K.}~\bibnamefont {Yabana}}, \bibinfo {author} {\bibfnamefont
  {L.}~\bibnamefont {Gallmann}}, \ and\ \bibinfo {author} {\bibfnamefont
  {U.}~\bibnamefont {Keller}},\ }\href {\doibase 10.1126/science.aag1268}
  {\bibfield  {journal} {\bibinfo  {journal} {Science}\ }\textbf {\bibinfo
  {volume} {353}},\ \bibinfo {pages} {916} (\bibinfo {year}
  {2016})}\BibitemShut {NoStop}%
\bibitem [{\citenamefont {Volkov}\ \emph {et~al.}(2019)\citenamefont {Volkov},
  \citenamefont {Sato}, \citenamefont {Schlaepfer}, \citenamefont {Kasmi},
  \citenamefont {Hartmann}, \citenamefont {Lucchini}, \citenamefont {Gallmann},
  \citenamefont {Rubio},\ and\ \citenamefont {Keller}}]{Volkov2019}%
  \BibitemOpen
  \bibfield  {author} {\bibinfo {author} {\bibfnamefont {M.}~\bibnamefont
  {Volkov}}, \bibinfo {author} {\bibfnamefont {S.~A.}\ \bibnamefont {Sato}},
  \bibinfo {author} {\bibfnamefont {F.}~\bibnamefont {Schlaepfer}}, \bibinfo
  {author} {\bibfnamefont {L.}~\bibnamefont {Kasmi}}, \bibinfo {author}
  {\bibfnamefont {N.}~\bibnamefont {Hartmann}}, \bibinfo {author}
  {\bibfnamefont {M.}~\bibnamefont {Lucchini}}, \bibinfo {author}
  {\bibfnamefont {L.}~\bibnamefont {Gallmann}}, \bibinfo {author}
  {\bibfnamefont {A.}~\bibnamefont {Rubio}}, \ and\ \bibinfo {author}
  {\bibfnamefont {U.}~\bibnamefont {Keller}},\ }\href {\doibase
  10.1038/s41567-019-0602-9} {\bibfield  {journal} {\bibinfo  {journal} {Nature
  Physics}\ }\textbf {\bibinfo {volume} {15}},\ \bibinfo {pages} {1145}
  (\bibinfo {year} {2019})}\BibitemShut {NoStop}%
\bibitem [{\citenamefont {Siegrist}\ \emph {et~al.}(2019)\citenamefont
  {Siegrist}, \citenamefont {Gessner}, \citenamefont {Ossiander}, \citenamefont
  {Denker}, \citenamefont {Chang}, \citenamefont {Schr{\"o}der}, \citenamefont
  {Guggenmos}, \citenamefont {Cui}, \citenamefont {Walowski}, \citenamefont
  {Martens}, \citenamefont {Dewhurst}, \citenamefont {Kleineberg},
  \citenamefont {M{\"u}nzenberg}, \citenamefont {Sharma},\ and\ \citenamefont
  {Schultze}}]{Siegrist2019}%
  \BibitemOpen
  \bibfield  {author} {\bibinfo {author} {\bibfnamefont {F.}~\bibnamefont
  {Siegrist}}, \bibinfo {author} {\bibfnamefont {J.~A.}\ \bibnamefont
  {Gessner}}, \bibinfo {author} {\bibfnamefont {M.}~\bibnamefont {Ossiander}},
  \bibinfo {author} {\bibfnamefont {C.}~\bibnamefont {Denker}}, \bibinfo
  {author} {\bibfnamefont {Y.-P.}\ \bibnamefont {Chang}}, \bibinfo {author}
  {\bibfnamefont {M.~C.}\ \bibnamefont {Schr{\"o}der}}, \bibinfo {author}
  {\bibfnamefont {A.}~\bibnamefont {Guggenmos}}, \bibinfo {author}
  {\bibfnamefont {Y.}~\bibnamefont {Cui}}, \bibinfo {author} {\bibfnamefont
  {J.}~\bibnamefont {Walowski}}, \bibinfo {author} {\bibfnamefont
  {U.}~\bibnamefont {Martens}}, \bibinfo {author} {\bibfnamefont {J.~K.}\
  \bibnamefont {Dewhurst}}, \bibinfo {author} {\bibfnamefont {U.}~\bibnamefont
  {Kleineberg}}, \bibinfo {author} {\bibfnamefont {M.}~\bibnamefont
  {M{\"u}nzenberg}}, \bibinfo {author} {\bibfnamefont {S.}~\bibnamefont
  {Sharma}}, \ and\ \bibinfo {author} {\bibfnamefont {M.}~\bibnamefont
  {Schultze}},\ }\href {\doibase 10.1038/s41586-019-1333-x} {\bibfield
  {journal} {\bibinfo  {journal} {Nature}\ }\textbf {\bibinfo {volume} {571}},\
  \bibinfo {pages} {240} (\bibinfo {year} {2019})}\BibitemShut {NoStop}%
\bibitem [{\citenamefont {Ghimire}\ \emph {et~al.}(2011)\citenamefont
  {Ghimire}, \citenamefont {DiChiara}, \citenamefont {Sistrunk}, \citenamefont
  {Agostini}, \citenamefont {DiMauro},\ and\ \citenamefont
  {Reis}}]{Ghimire2011}%
  \BibitemOpen
  \bibfield  {author} {\bibinfo {author} {\bibfnamefont {S.}~\bibnamefont
  {Ghimire}}, \bibinfo {author} {\bibfnamefont {A.~D.}\ \bibnamefont
  {DiChiara}}, \bibinfo {author} {\bibfnamefont {E.}~\bibnamefont {Sistrunk}},
  \bibinfo {author} {\bibfnamefont {P.}~\bibnamefont {Agostini}}, \bibinfo
  {author} {\bibfnamefont {L.~F.}\ \bibnamefont {DiMauro}}, \ and\ \bibinfo
  {author} {\bibfnamefont {D.~A.}\ \bibnamefont {Reis}},\ }\href {\doibase
  10.1038/nphys1847} {\bibfield  {journal} {\bibinfo  {journal} {Nature
  Physics}\ }\textbf {\bibinfo {volume} {7}},\ \bibinfo {pages} {138} (\bibinfo
  {year} {2011})}\BibitemShut {NoStop}%
\bibitem [{\citenamefont {Schubert}\ \emph {et~al.}(2014)\citenamefont
  {Schubert}, \citenamefont {Hohenleutner}, \citenamefont {Langer},
  \citenamefont {Urbanek}, \citenamefont {Lange}, \citenamefont {Huttner},
  \citenamefont {Golde}, \citenamefont {Meier}, \citenamefont {Kira},
  \citenamefont {Koch},\ and\ \citenamefont {Huber}}]{Schubert_2014_Sub}%
  \BibitemOpen
  \bibfield  {author} {\bibinfo {author} {\bibfnamefont {O.}~\bibnamefont
  {Schubert}}, \bibinfo {author} {\bibfnamefont {M.}~\bibnamefont
  {Hohenleutner}}, \bibinfo {author} {\bibfnamefont {F.}~\bibnamefont
  {Langer}}, \bibinfo {author} {\bibfnamefont {B.}~\bibnamefont {Urbanek}},
  \bibinfo {author} {\bibfnamefont {C.}~\bibnamefont {Lange}}, \bibinfo
  {author} {\bibfnamefont {U.}~\bibnamefont {Huttner}}, \bibinfo {author}
  {\bibfnamefont {D.}~\bibnamefont {Golde}}, \bibinfo {author} {\bibfnamefont
  {T.}~\bibnamefont {Meier}}, \bibinfo {author} {\bibfnamefont
  {M.}~\bibnamefont {Kira}}, \bibinfo {author} {\bibfnamefont {S.~W.}\
  \bibnamefont {Koch}}, \ and\ \bibinfo {author} {\bibfnamefont
  {R.}~\bibnamefont {Huber}},\ }\href {\doibase 10.1038/nphoton.2013.349}
  {\bibfield  {journal} {\bibinfo  {journal} {Nature Photonics}\ }\textbf
  {\bibinfo {volume} {8}},\ \bibinfo {pages} {119} (\bibinfo {year}
  {2014})}\BibitemShut {NoStop}%
\bibitem [{\citenamefont {Luu}\ \emph {et~al.}(2015)\citenamefont {Luu},
  \citenamefont {Garg}, \citenamefont {Kruchinin}, \citenamefont {Moulet},
  \citenamefont {Hassan},\ and\ \citenamefont {Goulielmakis}}]{Luu2015}%
  \BibitemOpen
  \bibfield  {author} {\bibinfo {author} {\bibfnamefont {T.~T.}\ \bibnamefont
  {Luu}}, \bibinfo {author} {\bibfnamefont {M.}~\bibnamefont {Garg}}, \bibinfo
  {author} {\bibfnamefont {S.~Y.}\ \bibnamefont {Kruchinin}}, \bibinfo {author}
  {\bibfnamefont {A.}~\bibnamefont {Moulet}}, \bibinfo {author} {\bibfnamefont
  {M.~T.}\ \bibnamefont {Hassan}}, \ and\ \bibinfo {author} {\bibfnamefont
  {E.}~\bibnamefont {Goulielmakis}},\ }\href {\doibase 10.1038/nature14456}
  {\bibfield  {journal} {\bibinfo  {journal} {Nature}\ }\textbf {\bibinfo
  {volume} {521}},\ \bibinfo {pages} {498} (\bibinfo {year}
  {2015})}\BibitemShut {NoStop}%
\bibitem [{\citenamefont {Tancogne-Dejean}\ \emph
  {et~al.}(2017{\natexlab{a}})\citenamefont {Tancogne-Dejean}, \citenamefont
  {M\"ucke}, \citenamefont {K\"artner},\ and\ \citenamefont
  {Rubio}}]{Nicolas_2017_Impact}%
  \BibitemOpen
  \bibfield  {author} {\bibinfo {author} {\bibfnamefont {N.}~\bibnamefont
  {Tancogne-Dejean}}, \bibinfo {author} {\bibfnamefont {O.~D.}\ \bibnamefont
  {M\"ucke}}, \bibinfo {author} {\bibfnamefont {F.~X.}\ \bibnamefont
  {K\"artner}}, \ and\ \bibinfo {author} {\bibfnamefont {A.}~\bibnamefont
  {Rubio}},\ }\href {\doibase 10.1103/PhysRevLett.118.087403} {\bibfield
  {journal} {\bibinfo  {journal} {Phys. Rev. Lett.}\ }\textbf {\bibinfo
  {volume} {118}},\ \bibinfo {pages} {087403} (\bibinfo {year}
  {2017}{\natexlab{a}})}\BibitemShut {NoStop}%
\bibitem [{\citenamefont {Yoshikawa}\ \emph {et~al.}(2017)\citenamefont
  {Yoshikawa}, \citenamefont {Tamaya},\ and\ \citenamefont
  {Tanaka}}]{Yoshikawa736}%
  \BibitemOpen
  \bibfield  {author} {\bibinfo {author} {\bibfnamefont {N.}~\bibnamefont
  {Yoshikawa}}, \bibinfo {author} {\bibfnamefont {T.}~\bibnamefont {Tamaya}}, \
  and\ \bibinfo {author} {\bibfnamefont {K.}~\bibnamefont {Tanaka}},\ }\href
  {\doibase 10.1126/science.aam8861} {\bibfield  {journal} {\bibinfo  {journal}
  {Science}\ }\textbf {\bibinfo {volume} {356}},\ \bibinfo {pages} {736}
  (\bibinfo {year} {2017})}\BibitemShut {NoStop}%
\bibitem [{\citenamefont {Tancogne-Dejean}\ \emph
  {et~al.}(2017{\natexlab{b}})\citenamefont {Tancogne-Dejean}, \citenamefont
  {M{\"u}cke}, \citenamefont {K{\"a}rtner},\ and\ \citenamefont
  {Rubio}}]{Tancogne-Dejean2017}%
  \BibitemOpen
  \bibfield  {author} {\bibinfo {author} {\bibfnamefont {N.}~\bibnamefont
  {Tancogne-Dejean}}, \bibinfo {author} {\bibfnamefont {O.~D.}\ \bibnamefont
  {M{\"u}cke}}, \bibinfo {author} {\bibfnamefont {F.~X.}\ \bibnamefont
  {K{\"a}rtner}}, \ and\ \bibinfo {author} {\bibfnamefont {A.}~\bibnamefont
  {Rubio}},\ }\href {\doibase 10.1038/s41467-017-00764-5} {\bibfield  {journal}
  {\bibinfo  {journal} {Nature Communications}\ }\textbf {\bibinfo {volume}
  {8}},\ \bibinfo {pages} {745} (\bibinfo {year}
  {2017}{\natexlab{b}})}\BibitemShut {NoStop}%
\bibitem [{\citenamefont {Liu}\ \emph {et~al.}(2017)\citenamefont {Liu},
  \citenamefont {Li}, \citenamefont {You}, \citenamefont {Ghimire},
  \citenamefont {Heinz},\ and\ \citenamefont {Reis}}]{Liu_2017_High}%
  \BibitemOpen
  \bibfield  {author} {\bibinfo {author} {\bibfnamefont {H.}~\bibnamefont
  {Liu}}, \bibinfo {author} {\bibfnamefont {Y.}~\bibnamefont {Li}}, \bibinfo
  {author} {\bibfnamefont {Y.~S.}\ \bibnamefont {You}}, \bibinfo {author}
  {\bibfnamefont {S.}~\bibnamefont {Ghimire}}, \bibinfo {author} {\bibfnamefont
  {T.~F.}\ \bibnamefont {Heinz}}, \ and\ \bibinfo {author} {\bibfnamefont
  {D.~A.}\ \bibnamefont {Reis}},\ }\href@noop {} {\bibfield  {journal}
  {\bibinfo  {journal} {Nature Physics}\ }\textbf {\bibinfo {volume} {13}},\
  \bibinfo {pages} {262} (\bibinfo {year} {2017})}\BibitemShut {NoStop}%
\bibitem [{\citenamefont {Tancogne-Dejean}\ and\ \citenamefont
  {Rubio}(2018)}]{Nicolas_2018_Atomic}%
  \BibitemOpen
  \bibfield  {author} {\bibinfo {author} {\bibfnamefont {N.}~\bibnamefont
  {Tancogne-Dejean}}\ and\ \bibinfo {author} {\bibfnamefont {A.}~\bibnamefont
  {Rubio}},\ }\href {\doibase 10.1126/sciadv.aao5207} {\bibfield  {journal}
  {\bibinfo  {journal} {Science Advances}\ }\textbf {\bibinfo {volume} {4}},\
  \bibinfo {pages} {eaao5207} (\bibinfo {year} {2018})}\BibitemShut {NoStop}%
\bibitem [{\citenamefont {Le~Breton}\ \emph {et~al.}(2018)\citenamefont
  {Le~Breton}, \citenamefont {Rubio},\ and\ \citenamefont
  {Tancogne-Dejean}}]{Guillaume_2018_High}%
  \BibitemOpen
  \bibfield  {author} {\bibinfo {author} {\bibfnamefont {G.}~\bibnamefont
  {Le~Breton}}, \bibinfo {author} {\bibfnamefont {A.}~\bibnamefont {Rubio}}, \
  and\ \bibinfo {author} {\bibfnamefont {N.}~\bibnamefont {Tancogne-Dejean}},\
  }\href {\doibase 10.1103/PhysRevB.98.165308} {\bibfield  {journal} {\bibinfo
  {journal} {Phys. Rev. B}\ }\textbf {\bibinfo {volume} {98}},\ \bibinfo
  {pages} {165308} (\bibinfo {year} {2018})}\BibitemShut {NoStop}%
\bibitem [{\citenamefont {Ghimire}\ and\ \citenamefont
  {Reis}(2019)}]{Ghimire2019}%
  \BibitemOpen
  \bibfield  {author} {\bibinfo {author} {\bibfnamefont {S.}~\bibnamefont
  {Ghimire}}\ and\ \bibinfo {author} {\bibfnamefont {D.~A.}\ \bibnamefont
  {Reis}},\ }\href {\doibase 10.1038/s41567-018-0315-5} {\bibfield  {journal}
  {\bibinfo  {journal} {Nature Physics}\ }\textbf {\bibinfo {volume} {15}},\
  \bibinfo {pages} {10} (\bibinfo {year} {2019})}\BibitemShut {NoStop}%
\bibitem [{\citenamefont {Kulander}\ \emph {et~al.}(1993)\citenamefont
  {Kulander}, \citenamefont {Schafer},\ and\ \citenamefont
  {Krause}}]{kulander1993dynamics}%
  \BibitemOpen
  \bibfield  {author} {\bibinfo {author} {\bibfnamefont {K.}~\bibnamefont
  {Kulander}}, \bibinfo {author} {\bibfnamefont {K.}~\bibnamefont {Schafer}}, \
  and\ \bibinfo {author} {\bibfnamefont {J.}~\bibnamefont {Krause}},\ }in\
  \href@noop {} {\emph {\bibinfo {booktitle} {Super-Intense Laser-Atom
  Physics}}}\ (\bibinfo  {publisher} {Springer},\ \bibinfo {year} {1993})\ pp.\
  \bibinfo {pages} {95--110}\BibitemShut {NoStop}%
\bibitem [{\citenamefont {Corkum}(1993)}]{Corkum_1994_Plasma}%
  \BibitemOpen
  \bibfield  {author} {\bibinfo {author} {\bibfnamefont {P.~B.}\ \bibnamefont
  {Corkum}},\ }\href {\doibase 10.1103/PhysRevLett.71.1994} {\bibfield
  {journal} {\bibinfo  {journal} {Phys. Rev. Lett.}\ }\textbf {\bibinfo
  {volume} {71}},\ \bibinfo {pages} {1994} (\bibinfo {year}
  {1993})}\BibitemShut {NoStop}%
\bibitem [{\citenamefont {Lewenstein}\ \emph
  {et~al.}(1994{\natexlab{b}})\citenamefont {Lewenstein}, \citenamefont
  {Balcou}, \citenamefont {Ivanov}, \citenamefont {L'Huillier},\ and\
  \citenamefont {Corkum}}]{PhysRevA.49.2117}%
  \BibitemOpen
  \bibfield  {author} {\bibinfo {author} {\bibfnamefont {M.}~\bibnamefont
  {Lewenstein}}, \bibinfo {author} {\bibfnamefont {P.}~\bibnamefont {Balcou}},
  \bibinfo {author} {\bibfnamefont {M.~Y.}\ \bibnamefont {Ivanov}}, \bibinfo
  {author} {\bibfnamefont {A.}~\bibnamefont {L'Huillier}}, \ and\ \bibinfo
  {author} {\bibfnamefont {P.~B.}\ \bibnamefont {Corkum}},\ }\href {\doibase
  10.1103/PhysRevA.49.2117} {\bibfield  {journal} {\bibinfo  {journal} {Phys.
  Rev. A}\ }\textbf {\bibinfo {volume} {49}},\ \bibinfo {pages} {2117}
  (\bibinfo {year} {1994}{\natexlab{b}})}\BibitemShut {NoStop}%
\bibitem [{\citenamefont {Vampa}\ \emph {et~al.}(2015)\citenamefont {Vampa},
  \citenamefont {McDonald}, \citenamefont {Orlando}, \citenamefont {Corkum},\
  and\ \citenamefont {Brabec}}]{Vampa_2015_semiclassical}%
  \BibitemOpen
  \bibfield  {author} {\bibinfo {author} {\bibfnamefont {G.}~\bibnamefont
  {Vampa}}, \bibinfo {author} {\bibfnamefont {C.~R.}\ \bibnamefont {McDonald}},
  \bibinfo {author} {\bibfnamefont {G.}~\bibnamefont {Orlando}}, \bibinfo
  {author} {\bibfnamefont {P.~B.}\ \bibnamefont {Corkum}}, \ and\ \bibinfo
  {author} {\bibfnamefont {T.}~\bibnamefont {Brabec}},\ }\href {\doibase
  10.1103/PhysRevB.91.064302} {\bibfield  {journal} {\bibinfo  {journal} {Phys.
  Rev. B}\ }\textbf {\bibinfo {volume} {91}},\ \bibinfo {pages} {064302}
  (\bibinfo {year} {2015})}\BibitemShut {NoStop}%
\bibitem [{\citenamefont {Ndabashimiye}\ \emph {et~al.}(2016)\citenamefont
  {Ndabashimiye}, \citenamefont {Ghimire}, \citenamefont {Wu}, \citenamefont
  {Browne}, \citenamefont {Schafer}, \citenamefont {Gaarde},\ and\
  \citenamefont {Reis}}]{Ndabashimiye_2016_Solid}%
  \BibitemOpen
  \bibfield  {author} {\bibinfo {author} {\bibfnamefont {G.}~\bibnamefont
  {Ndabashimiye}}, \bibinfo {author} {\bibfnamefont {S.}~\bibnamefont
  {Ghimire}}, \bibinfo {author} {\bibfnamefont {M.}~\bibnamefont {Wu}},
  \bibinfo {author} {\bibfnamefont {D.~A.}\ \bibnamefont {Browne}}, \bibinfo
  {author} {\bibfnamefont {K.~J.}\ \bibnamefont {Schafer}}, \bibinfo {author}
  {\bibfnamefont {M.~B.}\ \bibnamefont {Gaarde}}, \ and\ \bibinfo {author}
  {\bibfnamefont {D.~A.}\ \bibnamefont {Reis}},\ }\href {\doibase
  10.1038/nature17660} {\bibfield  {journal} {\bibinfo  {journal} {Nature}\
  }\textbf {\bibinfo {volume} {534}},\ \bibinfo {pages} {520} (\bibinfo {year}
  {2016})}\BibitemShut {NoStop}%
\bibitem [{\citenamefont {Ikemachi}\ \emph {et~al.}(2017)\citenamefont
  {Ikemachi}, \citenamefont {Shinohara}, \citenamefont {Sato}, \citenamefont
  {Yumoto}, \citenamefont {Kuwata-Gonokami},\ and\ \citenamefont
  {Ishikawa}}]{Ikemachi_2017_Trajectory}%
  \BibitemOpen
  \bibfield  {author} {\bibinfo {author} {\bibfnamefont {T.}~\bibnamefont
  {Ikemachi}}, \bibinfo {author} {\bibfnamefont {Y.}~\bibnamefont {Shinohara}},
  \bibinfo {author} {\bibfnamefont {T.}~\bibnamefont {Sato}}, \bibinfo {author}
  {\bibfnamefont {J.}~\bibnamefont {Yumoto}}, \bibinfo {author} {\bibfnamefont
  {M.}~\bibnamefont {Kuwata-Gonokami}}, \ and\ \bibinfo {author} {\bibfnamefont
  {K.~L.}\ \bibnamefont {Ishikawa}},\ }\href {\doibase
  10.1103/PhysRevA.95.043416} {\bibfield  {journal} {\bibinfo  {journal} {Phys.
  Rev. A}\ }\textbf {\bibinfo {volume} {95}},\ \bibinfo {pages} {043416}
  (\bibinfo {year} {2017})}\BibitemShut {NoStop}%
\bibitem [{\citenamefont {Kane}(1957)}]{Kane_1957_Band}%
  \BibitemOpen
  \bibfield  {author} {\bibinfo {author} {\bibfnamefont {E.~O.}\ \bibnamefont
  {Kane}},\ }\href {\doibase https://doi.org/10.1016/0022-3697(57)90013-6}
  {\bibfield  {journal} {\bibinfo  {journal} {Journal of Physics and Chemistry
  of Solids}\ }\textbf {\bibinfo {volume} {1}},\ \bibinfo {pages} {249 }
  (\bibinfo {year} {1957})}\BibitemShut {NoStop}%
\bibitem [{\citenamefont {Keldysh}(1965)}]{keldysh1965}%
  \BibitemOpen
  \bibfield  {author} {\bibinfo {author} {\bibfnamefont {L.}~\bibnamefont
  {Keldysh}},\ }\href@noop {} {\bibfield  {journal} {\bibinfo  {journal} {Sov.
  Phys. JETP}\ }\textbf {\bibinfo {volume} {20}},\ \bibinfo {pages} {1307}
  (\bibinfo {year} {1965})}\BibitemShut {NoStop}%
\bibitem [{\citenamefont {Balling}\ and\ \citenamefont
  {Schou}(2013)}]{balling2013femtosecond}%
  \BibitemOpen
  \bibfield  {author} {\bibinfo {author} {\bibfnamefont {P.}~\bibnamefont
  {Balling}}\ and\ \bibinfo {author} {\bibfnamefont {J.}~\bibnamefont
  {Schou}},\ }\href@noop {} {\bibfield  {journal} {\bibinfo  {journal} {Rep.
  Prog. Phys.}\ }\textbf {\bibinfo {volume} {76}},\ \bibinfo {pages} {036502}
  (\bibinfo {year} {2013})}\BibitemShut {NoStop}%
\bibitem [{\citenamefont {Tien}\ \emph {et~al.}(1999)\citenamefont {Tien},
  \citenamefont {Backus}, \citenamefont {Kapteyn}, \citenamefont {Murnane},\
  and\ \citenamefont {Mourou}}]{PhysRevLett.82.3883}%
  \BibitemOpen
  \bibfield  {author} {\bibinfo {author} {\bibfnamefont {A.-C.}\ \bibnamefont
  {Tien}}, \bibinfo {author} {\bibfnamefont {S.}~\bibnamefont {Backus}},
  \bibinfo {author} {\bibfnamefont {H.}~\bibnamefont {Kapteyn}}, \bibinfo
  {author} {\bibfnamefont {M.}~\bibnamefont {Murnane}}, \ and\ \bibinfo
  {author} {\bibfnamefont {G.}~\bibnamefont {Mourou}},\ }\href {\doibase
  10.1103/PhysRevLett.82.3883} {\bibfield  {journal} {\bibinfo  {journal}
  {Phys. Rev. Lett.}\ }\textbf {\bibinfo {volume} {82}},\ \bibinfo {pages}
  {3883} (\bibinfo {year} {1999})}\BibitemShut {NoStop}%
\bibitem [{\citenamefont {J\"urgens}\ \emph {et~al.}(2016)\citenamefont
  {J\"urgens}, \citenamefont {Jup\'e}, \citenamefont {Gyamfi},\ and\
  \citenamefont {Ristau}}]{doi:10.1117/12.2244833}%
  \BibitemOpen
  \bibfield  {author} {\bibinfo {author} {\bibfnamefont {P.}~\bibnamefont
  {J\"urgens}}, \bibinfo {author} {\bibfnamefont {M.}~\bibnamefont {Jup\'e}},
  \bibinfo {author} {\bibfnamefont {M.}~\bibnamefont {Gyamfi}}, \ and\ \bibinfo
  {author} {\bibfnamefont {D.}~\bibnamefont {Ristau}},\ }\href {\doibase
  10.1117/12.2244833} {\bibfield  {journal} {\bibinfo  {journal} {Proc.SPIE}\
  }\textbf {\bibinfo {volume} {10014}},\ \bibinfo {pages} {100141C} (\bibinfo
  {year} {2016})}\BibitemShut {NoStop}%
\bibitem [{\citenamefont {Chizhova}\ \emph {et~al.}(2017)\citenamefont
  {Chizhova}, \citenamefont {Libisch},\ and\ \citenamefont
  {Burgd\"orfer}}]{Chizhova_2017_High-harmonic}%
  \BibitemOpen
  \bibfield  {author} {\bibinfo {author} {\bibfnamefont {L.~A.}\ \bibnamefont
  {Chizhova}}, \bibinfo {author} {\bibfnamefont {F.}~\bibnamefont {Libisch}}, \
  and\ \bibinfo {author} {\bibfnamefont {J.}~\bibnamefont {Burgd\"orfer}},\
  }\href {\doibase 10.1103/PhysRevB.95.085436} {\bibfield  {journal} {\bibinfo
  {journal} {Phys. Rev. B}\ }\textbf {\bibinfo {volume} {95}},\ \bibinfo
  {pages} {085436} (\bibinfo {year} {2017})}\BibitemShut {NoStop}%
\bibitem [{\citenamefont {Yabana}\ and\ \citenamefont
  {Bertsch}(1996)}]{PhysRevB.54.4484}%
  \BibitemOpen
  \bibfield  {author} {\bibinfo {author} {\bibfnamefont {K.}~\bibnamefont
  {Yabana}}\ and\ \bibinfo {author} {\bibfnamefont {G.~F.}\ \bibnamefont
  {Bertsch}},\ }\href {\doibase 10.1103/PhysRevB.54.4484} {\bibfield  {journal}
  {\bibinfo  {journal} {Phys. Rev. B}\ }\textbf {\bibinfo {volume} {54}},\
  \bibinfo {pages} {4484} (\bibinfo {year} {1996})}\BibitemShut {NoStop}%
\end{thebibliography}%

\end{document}